\documentclass[reprint,prl,showpacs,superscriptaddress,longbibliography]{revtex4-1}
\usepackage{amsmath,amssymb,bm,mathrsfs,graphicx, braket, times,color}
\usepackage[colorlinks=true,citecolor=blue,linkcolor=blue]{hyperref}
\newcommand{\sg}[1]{{\bf\em{#1}}}
\usepackage{soul}
\usepackage{xcolor}

\begin{document}

\title{Space group theory of photonic bands}

\author{Haruki Watanabe} 
\affiliation{Department of Applied Physics, University of Tokyo, Tokyo 113-8656, Japan}

\author{Ling Lu}
\affiliation{Institute of Physics, Chinese Academy of Sciences/Beijing National
Laboratory for Condensed Matter Physics, Beijing, China}

\begin{abstract}
The wide-range application of photonic crystals and metamaterials benefits from the enormous design space of three-dimensional sub-wavelength structures. 
In this work, we study the space group constraints on photonic dispersions for all 230 space groups with time-reversal symmetry. 
Our theory carefully treats the unique singular point of photonic bands at zero frequency and momentum, which distinguishes photonic bands from their electronic counterpart. The results are given in terms of minimal band connectivities at zero~($M$) and non-zero frequencies~($M'$).
Topological band degeneracies are guaranteed to be found in space groups that do not allow band gaps between the second and third photonic bands~($M>2$).
Our work provides theoretical guidelines for the choice of spatial symmetries in photonics design.
\end{abstract}

\maketitle

\begin{figure}[t]
\includegraphics[width=0.50 \textwidth]{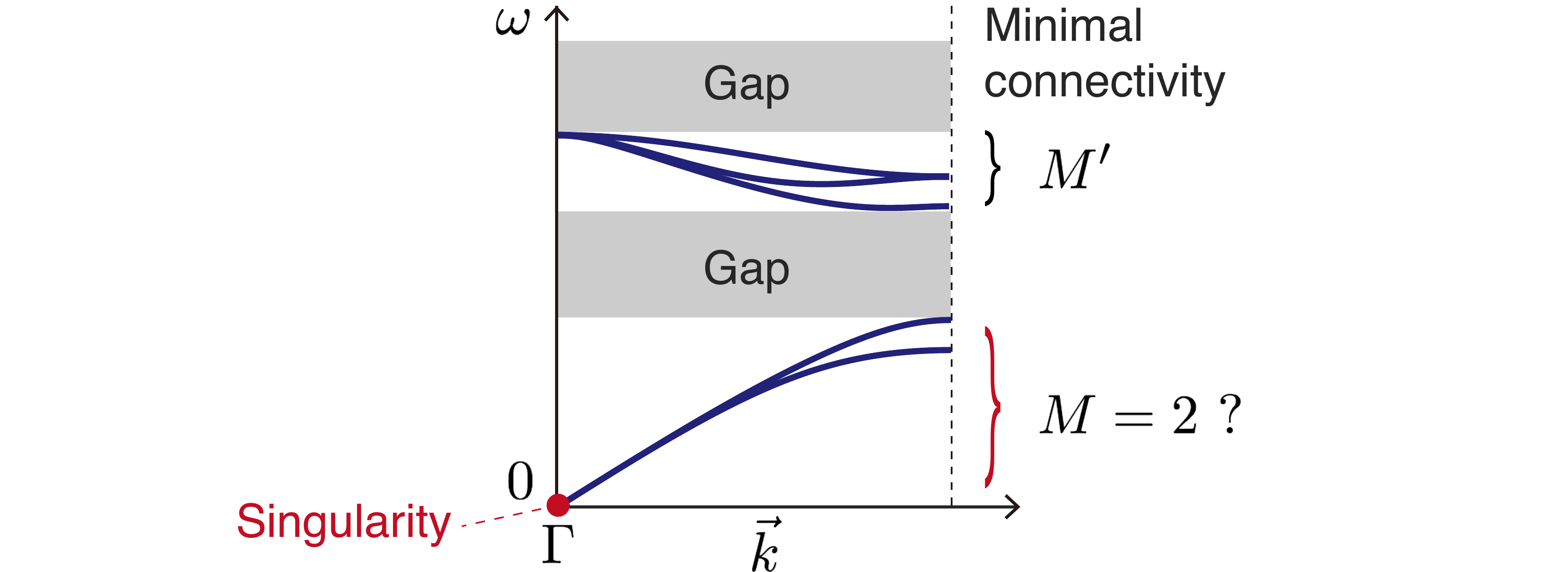}
\caption{Two types of band connectivities ($M$ and $M'$) are distinguished by whether they are connected to $\omega = |\vec{k}| = 0$.  
\label{fig1}}
\end{figure}

\paragraph{Introduction.}--- A photonic system with translational symmetry is described by a band structure showing the frequency spectrum as a function of the lattice momentum. The appearance of band gaps, where the density of states vanishes, is the most prominent feature in photonic band structures.  How bands connect to each other over the Brillouin zone and whether band gaps can open at specific frequency levels are highly constrained by the symmetry of the underlying lattice.
Similar symmetry constraints on electronic band structures were recently studied for space groups~\cite{usPRL,NC,Bradlyn17}. However, these results are only translatable to nonzero-frequency bands of photonics. The standard treatment fails for photonic crystals because their band structure has an intrinsic singularity at zero frequency and momentum, as illustrated in Fig.~\ref{fig1}. Systematic understanding of whether these two gapless bands can be separated from the higher-frequency bands is important for constructing photonic crystals with targeted properties. Low-frequency bands also have high priority in practice because their features are quite forgiving for fabrication imperfections. In this work, we develop a group theoretic approach for photonic bands and determine the possible gap positions and band connectivities of time-reversal-invariant photonic crystals for all 230 space groups.  Our results will guide the choice of space groups in designing photonic crystals, metamaterials, and topological photonic lattices.

\paragraph{Motivations.}--- The study of photonic crystals began with the search for three-dimensional band gaps~\cite{yablonovitch1987inhibited,john1987strong,Joannopoulos}. The first complete gap was discovered between the second and third bands in the diamond lattice~(space group \sg{227})~\cite{Ho}, and it has remained the largest gap in dielectric photonic crystals ever since~\cite{maldovan2003exploring,maldovan2004diamond,michielsen2003photonic,men2014robust}. (Hereafter, we refer to a space group by its number assigned in Ref.~\cite{ITC} in bold italic font.) It is reasonable that the largest gap opens between the lowest bands, where the density of states is the lowest.
We find that \sg{227} is actually the largest space group that allows separation between the second and third bands. This justifies why no larger gaps than that of the diamond lattice have been found using dielectrics.

Metamaterials of periodic metal composites~\cite{cai2010optical} can be understood as metallic photonic crystals~(or plasmonic crystals)~\cite{raman2010photonic,raman2013metamaterial}. Some of our results also apply to them, as we discuss later.

Topological photonics started by realizing that photonic band structures could be distinct in their global configurations of wavefunctions below a band gap~\cite{lu2014topological,ozawa2018topological}. Therefore, knowing the general condition of band gaps is a prerequisite in the search for topological photonic bands. 
Our results show space groups in which topological band degeneracies, such as Weyl points~\cite{lu2013weyl} and nodal lines, can be found between the second band and higher bands.  

\paragraph{Minimal connectivity: $M$ and $M'$.}--- Minimal band connectivities $M$ and $M'$, illustrated in Fig.~\ref{fig1}, are the key quantity to be determined in this paper. They represent the minimal number of bands that are required to connect with each other somewhere in the Brillouin zone as a result of the spatial symmetry and the time-reversal symmetry~\cite{michel1999connectivity}. $M$ denotes the set of bands connected to $\omega = |\vec{k}| = 0$. Obviously, $M\geqslant2$ for dielectric photonic crystals because of the existence of the two gapless modes. $M'\geqslant1$ denotes the rest of the bands with higher frequencies.  Throughout the paper, a band gap means that the two bands below and above do not touch at any momentum in the entire Brillouin zone.

We obtain $M$ and $M'$ for each space group by examining its compatibility relations and making use of its sub/supergroup relations. These principles are valid regardless of the details of the system, such as unit-cell shapes and material dispersions in permittivity or permeability.
The compatibility relations, reviewed in our Supplementary Material (SM), are relations among symmetry representations at high-symmetry momenta. They require that the combination of representations used in the band structure is globally consistent over the entire Brillouin zone. $M$ and $M'$ are the minimal number of bands that satisfy the compatibility relations. 

A translationengleiche-($t$-)subgroup~\cite{ITC} is a subgroup in which the translation symmetry is preserved while other group elements are removed; $t$-supergroup is also defined likewise. The identical translation implies the same Brillouin zone structure. Intuitively, a higher-symmetry~($t$-supergroup) implies a larger band connectivity and, conversely, a lower-symmetry~($t$-subgroup) implies a smaller band connectivity. For example, if a space group allows $M=2$, all of its $t$-subgroups have $M=2$, while all of its $t$-supergroups have a connectivity $\geqslant{M}$.

\paragraph{Space-group constrains on $M'$.}--- Recently, a similar approach has been applied to electronic band structures, where the possible connectivities of bands are determined for all space groups~\cite{usPRL,NC,Bradlyn17}. Our $M'$ values can be readily obtained by halving the values in Ref.~\cite{usPRL} (labeled $\nu$ in Table S2 of Ref.~\cite{usPRL}) to take into account the lack of spin in the photonic problem.
We show the results in Table~S1 in the SM.
For symmorphic space groups~(containing neither screw nor glide), $M'=1$ because of the existence of 1D representations. For nonsymmorphic space groups, $M'\geqslant2$, as we explain below. This table applies to any bosonic band structure.

It is important to note that $M\neq M'$ in general.
For example, the single gyroid~(belonging to space group \sg{214}) has a band gap between the second and third bands (i.e., $M=2$)~\cite{lu2013weyl}. However, according to Table~S1, $M'=4$ for this space group.  
This apparent mismatch motivated us to revisit the band theory for photonic crystals. The key difference lies at the singularity at $\omega = |\vec{k}| = 0$.

\paragraph{Singularity at zero.}--- Maxwell's equations in free space do not have a converging eigen-solution exactly at $\omega = |\vec{k}| = 0$.
Around this point, electric and magnetic fields for a wavevector~$\vec{k}$ polarize in the plane perpendicular to $\vec{k}$ and do not converge as $\vec{k}\rightarrow\vec{0}$. Therefore, wavefunctions at this singular point may not form a representation of the symmetry of the system, and the standard band theory and group theory fail in general. In comparison, electronic band structures do not have such singularities. Phonons also disperse linearly at zero, but they can have a converging solution with three branches representing a vector in three dimensions~\cite{walker1995site}, in contrast with photons, which have only two transverse gapless branches.

Fortunately, anywhere away from the singular point, wavefunctions are smooth and representations can be assigned. 
We argue that, in dielectric photonic crystals, two gapless modes transform in the same way as planewaves under spatial symmetries around the singular point. This can be understood through a “thought experiment”, in which we adiabatically increase the dielectric constant from the free-space unity~($\epsilon=1$) to any values of $\epsilon\geqslant1$, while maintaining the assumed spatial symmetries. In this adiabatic process, the dispersion curves can move, but their symmetry representations never change.  Consequently, the two gapless photon dispersions have the same symmetry eigenvalues as the planewaves of uniform vectorial electric fields~(or pseudo-vectorial magnetic fields). They are $e^{i\frac{2\pi}{n}}$ and $e^{-i\frac{2\pi}{n}}$ for an $n$-fold rotation and $+1$ and $-1$ for a mirror along each high-symmetry line around $\omega = |\vec{k}| = 0$. We use these rules to study $M$.

\paragraph{Metallic photonic crystals.}--- 
The properties of the lowest-frequency bands of some metallic photonic crystals are different than those for dielectric crystals. Here, we classify metallic photonic crystals into three classes according to their low-frequency dispersions. Our results in this work apply to the first two classes, but \emph{not} the third one.

The first class of metallic photonic crystals has the same low-frequency dispersion as dielectrics.  An example is given by a periodic arrangement of isolated metallic elements that are disconnected from each other in every spatial direction. Our results for $M$ and $M'$ clearly apply to this class.  The second class has a band gap toward the zero frequency, which can be interpreted as $M = 0$. Our $M'$ still applies in this case. Examples include a single metallic network connected in all three dimensions. The third class, discovered recently~\cite{shin2007three,chen2018metamaterials,smith2018violating}, has exotic low-frequency bands. Although our results for $M'$ still apply, the $M$ values of this class require a separate treatment.

\begin{figure}[t]
\includegraphics[width=0.5 \textwidth]{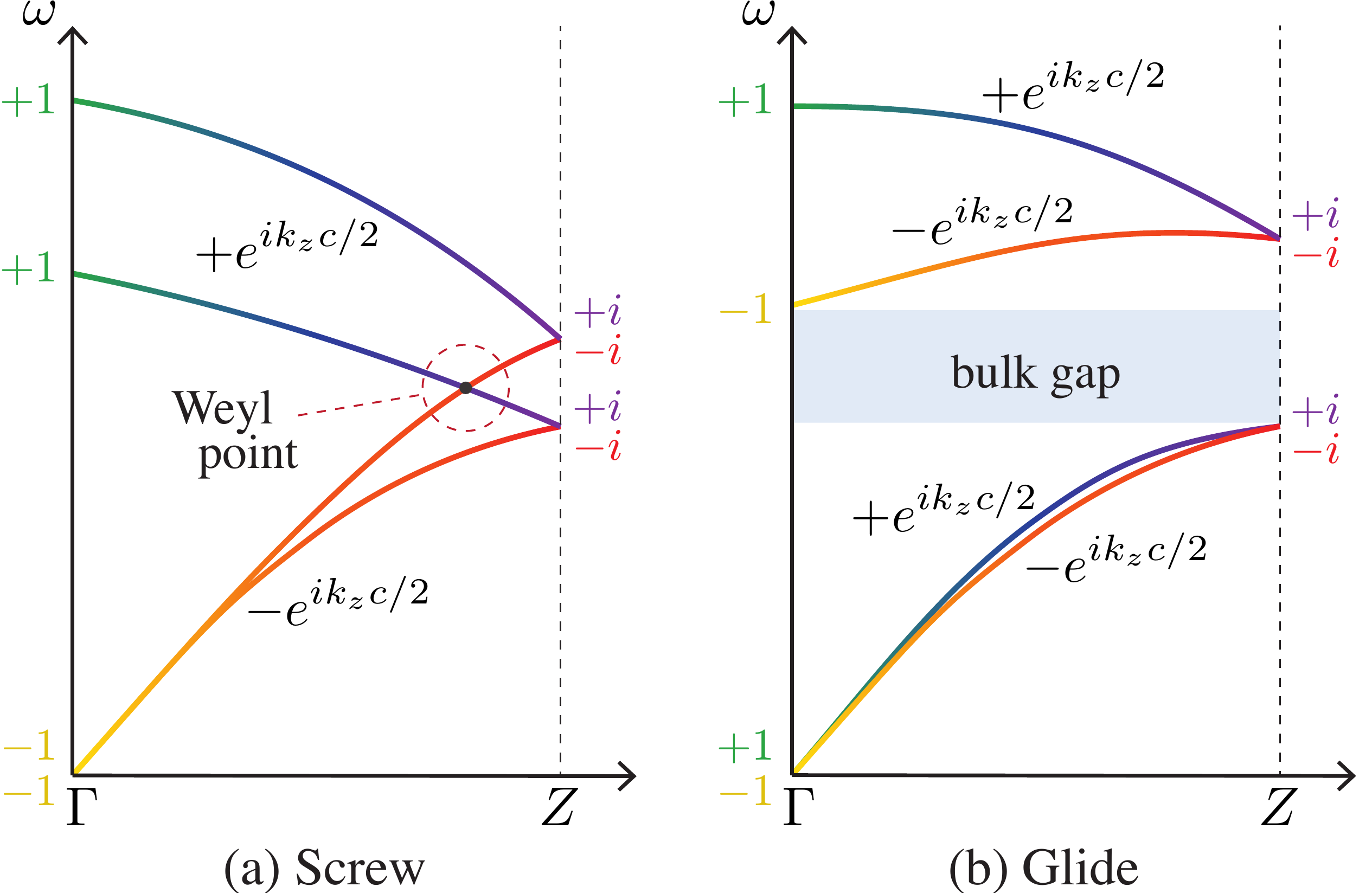}
\caption{Consequence of a nonsymmorphic symmetry on band connectivities. (a) $M=4$ for \sg{4}~($P2_1$) containing a two-fold screw. (b) $M=2$ for \sg{7}~($Pc$) containing a glide. The numbers aside the vertical axes and the colors of the dispersions indicate the eigenvalue of the nonsymmorphic symmetry.
\label{fig2}}
\end{figure}

\paragraph{Nonsymmorphic examples of $M>2$.}---
Below, we discuss two sets of examples~(nonsymmorphic and symmorphic space groups) to explain how the above photonic band theory is applied to derive $M$ for 230 space groups.  

First, we demonstrate that a two-fold screw symmetry protects crossing of the lowest four bands somewhere along a high-symmetry line (i.e., $M=4$), while a glide symmetry does not.
To this end, let us start with reviewing the basics of a two-fold screw symmetry. Space group \sg{4}~($P2_1$) is generated by translations and a two-fold screw rotation $S_{2z}$ that maps $(x,y,z)$ to $(-x,-y,z+\frac{c}{2})$ ($a$, $b$, and $c$ are the lattice constants). The line $\Gamma$-$Z$ connecting $\Gamma=(0,0,0)$ and $Z=(0,0,\frac{\pi}{c})$ is invariant under the screw operation.  Because $(S_{2z})^2=T_z$ is the unit lattice translation in $z$, the eigenvalues of $S_{2z}$ are $\pm e^{i k_zc/2}$. The factor of $\frac{1}{2}$ in the exponent implies that the two eigenvalues interchange when $k_z$ increases by $\frac{2\pi}{c}$. As a result, a branch with an eigenvalue of $+ e^{i k_zc/2}$ must cross with another branch with $-e^{i k_zc/2}$ somewhere along this line. Therefore, the band connectivity~($M$ and $M'$) must always be even. In the presence of time-reversal symmetry, the crossing point is pinned to the $Z$ point, where the $S_z$-eigenvalues are purely imaginary and form a pair under time-reversal symmetry.

Now, recall that the fields around $\Gamma$ transform in the same way as planewaves. Thus, they flip signs under the $\pi$-rotation part $C_{2z}: (x,y,z)\mapsto(-x,-y,z)$ of the screw. This implies that both of the gapless branches have the screw eigenvalue, $-e^{i k_zc/2}$. As a result, in total, four bands have to cross each other, resulting in a linear crossing between the second and third bands, as shown in Fig.~\ref{fig2}(a). This conclusion holds in any space group that contains \sg{4} as a $t$-subgroup. The crossing becomes a nodal line when a $t$-supergroup contains the inversion symmetry; otherwise, it is a Weyl point, as in \sg{4}.

Let us compare this result for a two-fold screw with a glide reflection symmetry.  
Although a glide and a two-fold screw usually result in the same band connectivity in electronic band structures, the effect is clearly different in photonic bands. To see this, let $G_z$ be the glide operation transforming $(x,y,z)$ to $(x,-y,z+\frac{c}{2})$. Eigenvalues of $G_z$ along the line $\Gamma$-$Z$ are $\pm e^{i k_zc/2}$, the same as $S_{2z}$, as $(G_z)^2=T_z=e^{ik_z c}$. However, only one of the two gapless photons flips sign under the mirror part  $M_{z}: (x,y,z)\mapsto(x,y,-z)$ of $G_z$, unlike the two-fold rotation part of $S_{2z}$. 
As a consequence, there is a consistent assignment of glide eigenvalues with only two bands, as shown in Fig.~\ref{fig2}(b). Therefore, $M=2$ space group  \sg{7} ($Pc$), generated by translations and the glide symmetry.

\begin{figure}[t]
\includegraphics[width=0.5 \textwidth]{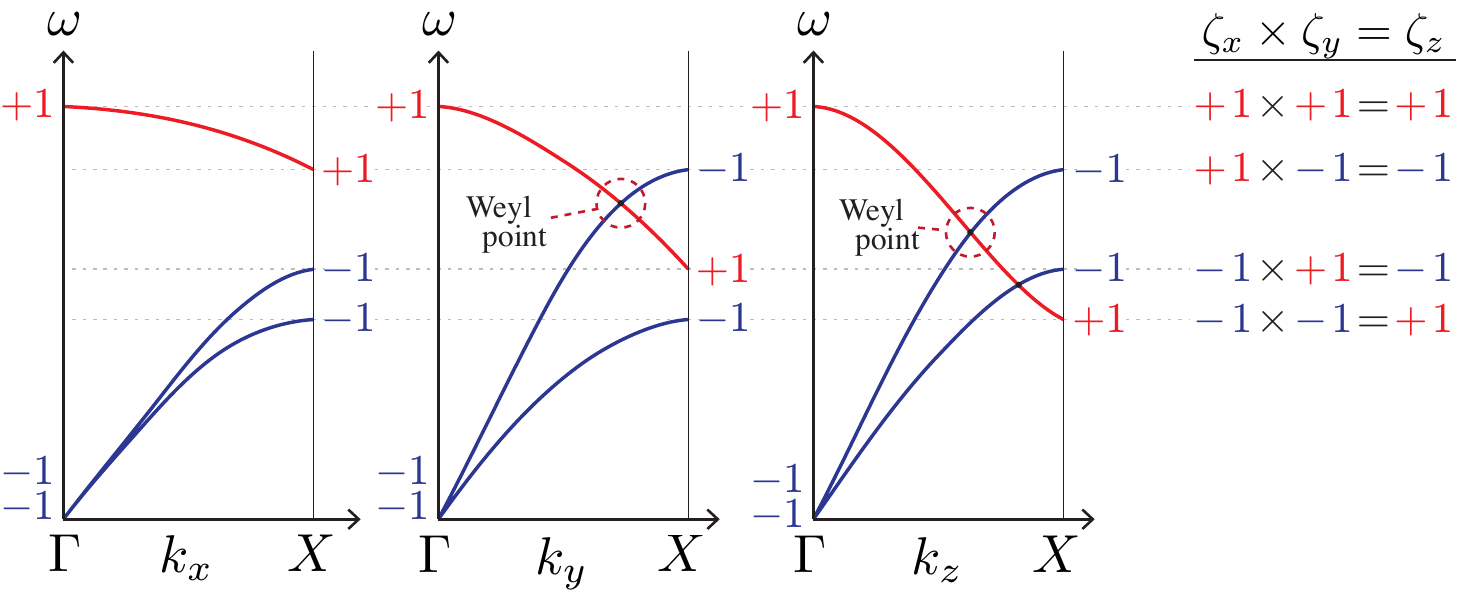}
\caption{A possible band structure of space group \sg{23}~($I222$) along three different lines connecting $\Gamma$ and $X$.   The numbers aside the vertical axes indicate the eigenvalues $\zeta_\alpha$ of the $C_{2\alpha}$ rotation for $\alpha=x,y,z$.  The three eigenvalues at the same point~$X$ must satisfy $\zeta_x\zeta_y=\zeta_z$ except at $\Gamma$.
\label{fig3}}
\end{figure}

\begin{figure*}[t]
\includegraphics[width= \textwidth]{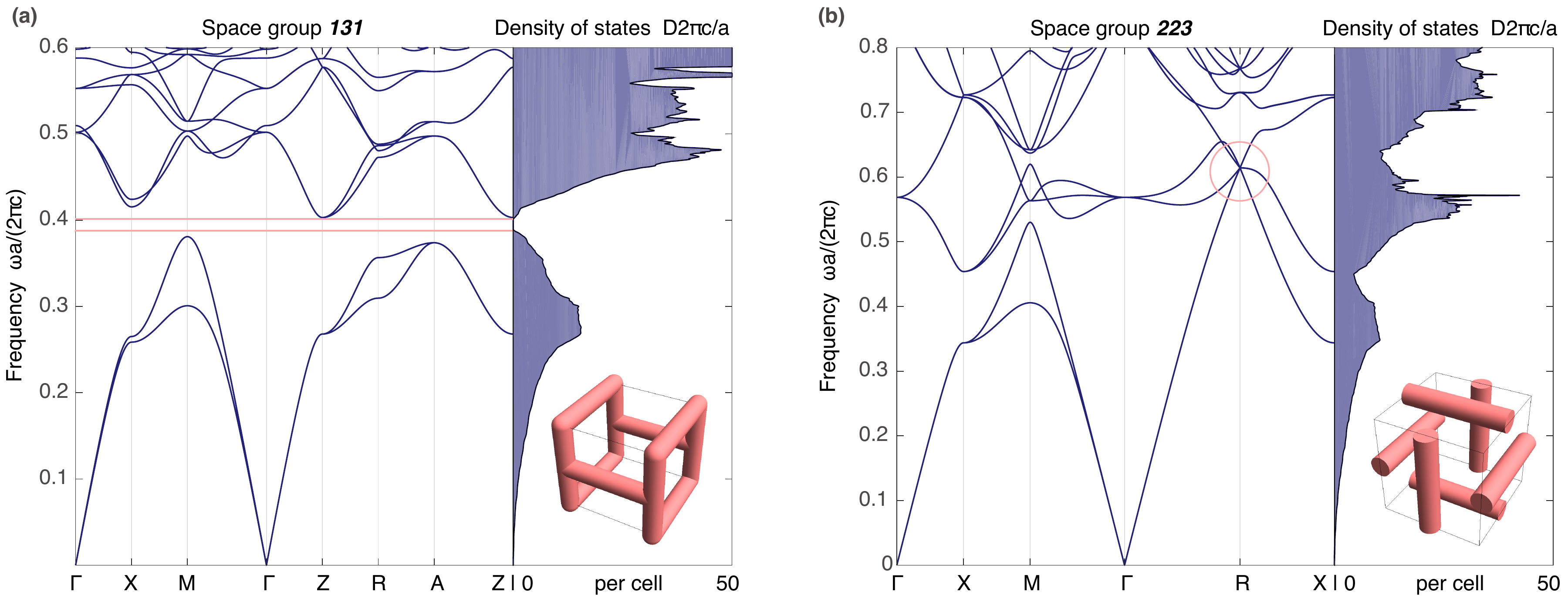}
\caption{Examples of new photonic crystals with $M=2$ and $M>2$.
(a) Space group \sg{131}~($P4_2/mmc$) with rod radius 0.15$a$. There is a full band gap of 5\% between the second the third bands. All bands along $A$-$Z$ are doubly-degenerate.
(b) Space group \sg{223}~($Pm\bar{3}n$) with rod radius 0.1$a$. There is a six-fold degeneracy point at $R$.  All bands along $R$-$X$ are doubly-degenerate.
Insets are the real-space structures of dielectric constant $\epsilon=13$ in the cubic unit cell of lattice constant $a$.
}
\label{fig4}
\end{figure*}

\paragraph{Symmorphic examples of $M>2$.}--- Next, let us demonstrate that a band gap between the second and third bands can be prohibited, even for symmorphic space groups.
We discuss space group \sg{23} as an example and show that it has $M=3$. Space group \sg{23}~($I222$) has three $\pi$-rotations $C_{2\alpha}$ about $\alpha=x$, $y$, and $z$ axes, in addition to the lattice translations defined by the primitive lattice vectors $\vec{a}_1=\frac{1}{2}(-a,b,c)$, $\vec{a}_2=\frac{1}{2}(a,-b,c)$, and $\vec{a}_3=\frac{1}{2}(a,b,-c)$. The corresponding reciprocal lattice vectors are $\vec{b}_1=(0,\frac{2\pi}{b},\frac{2\pi}{c})$, $\vec{b}_2=(\frac{2\pi}{a},0,\frac{2\pi}{c})$, and $\vec{b}_3=(\frac{2\pi}{a},\frac{2\pi}{b},0)$. Because $(C_{2\alpha})^2=1$, the eigenvalues $\zeta_\alpha=\pm 1$ of $C_{2\alpha}$ do not depend on $\vec{k}$, unlike the nonsymmorphic symmetries discussed above.  

Let us first focus on the line connecting $\Gamma$ to $X_1=(\frac{2\pi}{a},0,0)$, which is symmetric under $C_{2x}$. The two gapless dispersions have the $-1$ eigenvalues of $C_{2x}$ on the entire line.  Similarly, the line $\Gamma$-$X_2$~[$X_2=(0,\frac{2\pi}{b},0)$] has the $C_{2y}$ symmetry and $-1$ eigenvalues for the two gapless dispersions. The line connecting $\Gamma$ to $X_3=(0,0,\frac{2\pi}{c})$ is also similar. Now, note that $X_1$, $X_2$, and $X_3$ are actually identical points in the Brillouin zone, as $X_1-X_2=\vec{b}_2-\vec{b}_1$, for instance, is a reciprocal lattice vector. We call this point $X$~($=X_1=X_2=X_3$), and there are three inequivalent lines connecting $\Gamma$ to $X$. As a result, the $X$ point has all three $\pi$-rotations with the multiplication rule $C_{2x}C_{2y}=C_{2z}$.
However, the two gapless photons both have the eigenvalues $\zeta_x=\zeta_y=\zeta_z=-1$ and cannot fulfill $\zeta_z=\zeta_x\zeta_y$ by themselves. Therefore, there must be at least one extra band supplying a $+1$ eigenvalue; Fig.~\ref{fig3} shows one such possibility.

In this \sg{23} example, $X_{1}=X_{2}=X_{3}$ is a property of the body-centered lattice.
In contrast, other symmorphic space groups with the identical point group but with different lattice translations do not share the same conclusion.
For example, \sg{22}~($F222$) of the face-centered lattice, \sg{21}~($C222$) of the base-centered lattice, and \sg{16}~($P222$) of the primitive lattice all have $M=2$ (see the following list).

\paragraph{Space groups possible with $M=2$.} --- Applying a similar analysis, we determine 104 space groups that allow a band gap of $M=2$:
\begin{quote}
\sg{1}, \sg{2}, \sg{3}, \sg{5}, \sg{6}, \sg{7}, \sg{8}, \sg{9}, \sg{10}, \sg{12}, \sg{13}, \sg{15}, \sg{16}, \sg{21}, \sg{22}, \sg{24}, \sg{25}, \sg{27}, \sg{28},
\sg{30}, \sg{32}, \sg{34}, \sg{35}, \sg{37}, \sg{38}, \sg{\underline{39}}, \sg{40}, \sg{41}, \sg{42}, \sg{43}, \sg{44}, \sg{45},
 \sg{46}, \sg{47}, \sg{\underline{48}}, \sg{\underline{49}}, \sg{\underline{50}}, \sg{65}, \sg{66}, \sg{\underline{68}}, \sg{70}, \sg{74}, \sg{75}, \sg{77}, \sg{79}, \sg{80}, \sg{81}, \sg{82}, \sg{83}, \sg{84}, \sg{\underline{86}}, \sg{88},
\sg{89}, \sg{93}, \sg{98}, \sg{99}, \sg{\underline{100}}, \sg{\underline{101}}, \sg{\underline{102}}, \sg{105},
 \sg{\underline{107}}, \sg{\underline{108}}, \sg{109}, \sg{111}, \sg{112}, \sg{115}, \sg{119}, \sg{122}, \sg{123}, \sg{\underline{131}}, \sg{141}, \sg{143}, \sg{146}, \sg{147}, \sg{148}, \sg{149}, \sg{150}, \sg{155}, \sg{156},
\sg{157}, \sg{160}, \sg{162}, 
\sg{164}, \sg{166}, \sg{168}, \sg{174}, \sg{175}, \sg{177}, \sg{183}, \sg{187}, \sg{189}, \sg{\underline{191}}, \sg{195}, \sg{196}, \sg{199}, \sg{200}, \sg{203}, \sg{207}, \sg{210}, \sg{\underline{214}}, \sg{215}, \sg{216}, \sg{\underline{221}}, \sg{\underline{227}}.
\end{quote}
In the SM, we derive the solution to the compatibility relations for $M=2$ for the 16 key groups underscored in the above list. 
All of these space groups here are a $t$-subgroup of at least one of the 16 key space groups, as summarized in Table~S2, which implies $M=2$.

Our results are consistent with the known photonic band gaps between the second and third bands.
For example, the three highest space groups of $M=2$ are diamond~(\sg{227})~\cite{Ho}, simple cubic~(\sg{221})~\cite{sozuer1993photonic}, and single gyroid~(\sg{214})~\cite{lu2013weyl}.
To further verify our prediction, we provide a new example of \sg{131} with $M=2$. Its band structure and density of states~\cite{liu2018generalized} are plotted in Fig.~\ref{fig4}(a), showing a full band gap between the lowest dielectric bands.

\paragraph{Space groups of $M>2$.}--- One can also show that a band gap between the second and the third bands is not allowed by the compatibility relations for the other 126 space groups. 
All of these space groups are $t$-supergroups of at least one of the following 22 key space groups:
\begin{quote}
\sg{4}, \sg{23}, \sg{67}, \sg{69}, \sg{73}, \sg{85}, \sg{87}, \sg{103}, \sg{104}, \sg{106}, \sg{110}, \sg{116}, \sg{117}, \sg{118}, \sg{120}, \sg{144}, \sg{145}, \sg{158}, \sg{159}, \sg{161}, \sg{201}, \sg{208}.
\end{quote}
We explain why it is not possible to realize $M=2$ for these key space groups one by one in Sec.~V in the SM.

Our results are consistent with the band structures of known photonic crystals in which no band gaps are found between the second and third bands, such as the hexagonal close packing~(\sg{194})~\cite{busch1998photonic},  tetrahedral~(\sg{224})~\cite{yang2017weyl}, face-centered-cubic~(\sg{225})~\cite{sozuer1992photonic}, body-centered cubic~(\sg{229})~\cite{luo2002all}, and double-gyroid~(\sg{230})~\cite{lu2013weyl} dielectric photonic crystals.

To further verify our prediction, we provide a new example of \sg{223} with $M>2$. In the example shown in Fig.~\ref{fig4}, the lowest two bands are a part of a six-dimensional representation at $R$~\cite{Bradley}. This is a new type of topological band-crossing point beyond the Weyl and Dirac points~\cite{bradlyn2016beyond,fang2017diagnosis,shiozaki2018atiyah}.  Gapping these topological degeneracies by lowering the symmetry can generate topological band gaps and interfacial states in three dimensions~\cite{lu2016symmetry,lu2016topological,ochiai2017gapless}.

\paragraph{Outlook.}---
This work presents a systematic symmetry analysis of photonic bands for 230 space groups, which has been lacking since the discovery of photonic crystals. The results in Tables~S1 ($M'$), S2 ($M=2$), and S3 ($M>2$) in the SM provide useful design insights for photonic crystals, metamaterials, and topological lattices.
Future studies are expected toward a more exhaustive knowledge of space-group constraints on photonic bands.
First, satisfying the compatibility relations is only a necessary condition for band gaps, as it only concerns the representations of high-symmetry momentum points, lines, or surfaces. Topological degeneracies of Weyl points and nodal lines can take place at general momenta. Second, $M>2$ values can be further pinpointed for all 126 space groups with $M\neq2$. Third, instead of the minimal connectivity, all intrinsic band connectivity values can be worked out.
Last, the minimal band connectivities of metallic photonic crystals with irregular zero-frequency bands~\cite{shin2007three,chen2018metamaterials,smith2018violating}, the \emph{only} case where our current results do not apply, could be determined by extending the analysis in this work.

\begin{acknowledgements}
We thank Chen Fang, Hoi Chun Po, Qinghui Yan and Hengbin Cheng for useful discussions.
H.W. is supported by JSPS KAKENHI Grant Numbers JP17K17678.
L.L. was supported by the National key R\&D Program of China under Grant No.~2017YFA0303800, 2016YFA0302400 and by NSFC under Project No.~11721404.  
\end{acknowledgements}
\bibliography{references}

\clearpage
\onecolumngrid

\renewcommand{\thefigure}{S\arabic{figure}}
\renewcommand{\thetable}{S\arabic{table}}


\section{Minimal connectivity of nonzero-frequency bands~($M'$) of 230 space groups}
\begin{table*}[h]
\begin{center}
\caption{
Minimal band connectivity~($M'$), of nonzero-frequency bands, of a time-reversal-invariant photonic crystal~(both dielectric or metallic) for all 230 space groups.
We listed $M'$ for all 157 nonsymmorphic groups. Those not listed are symmorphic space group whose $M'=1$.
\label{tabM'}
}
\begin{tabular}{|cc|cc|cc|cc|cc|cc|cc|}
No.	&$M'$	&No.	&$M'$	&No.	&$M'$	&No.	&$M'$	&No.	&$M'$	&No.	&$M'$	&No.	&$M'$	\\\hline
\sg{4}&$2$&\sg{39}&$2$&\sg{66}&$2$&\sg{100}&$2$&\sg{129}&$2$&\sg{165}&$2$&\sg{201}&$2$\\
\sg{7}&$2$&\sg{40}&$2$&\sg{67}&$2$&\sg{101}&$2$&\sg{130}&$4$&\sg{167}&$2$&\sg{203}&$2$\\
\sg{9}&$2$&\sg{41}&$2$&\sg{68}&$2$&\sg{102}&$2$&\sg{131}&$2$&\sg{169}&$6$&\sg{205}&$4$\\
\sg{11}&$2$&\sg{43}&$2$&\sg{70}&$2$&\sg{103}&$2$&\sg{132}&$2$&\sg{170}&$6$&\sg{206}&$4$\\
\sg{13}&$2$&\sg{45}&$2$&\sg{72}&$2$&\sg{104}&$2$&\sg{133}&$4$&\sg{171}&$3$&\sg{208}&$2$\\
\sg{14}&$2$&\sg{46}&$2$&\sg{73}&$4$&\sg{105}&$2$&\sg{134}&$2$&\sg{172}&$3$&\sg{210}&$2$\\
\sg{15}&$2$&\sg{48}&$2$&\sg{74}&$2$&\sg{106}&$4$&\sg{135}&$4$&\sg{173}&$2$&\sg{212}&$4$\\
\sg{17}&$2$&\sg{49}&$2$&\sg{76}&$4$&\sg{108}&$2$&\sg{136}&$2$&\sg{176}&$2$&\sg{213}&$4$\\
\sg{18}&$2$&\sg{50}&$2$&\sg{77}&$2$&\sg{109}&$2$&\sg{137}&$2$&\sg{178}&$6$&\sg{214}&$4$\\
\sg{19}&$4$&\sg{51}&$2$&\sg{78}&$4$&\sg{110}&$4$&\sg{138}&$4$&\sg{179}&$6$&\sg{218}&$2$\\
\sg{20}&$2$&\sg{52}&$4$&\sg{80}&$2$&\sg{112}&$2$&\sg{140}&$2$&\sg{180}&$3$&\sg{219}&$2$\\
\sg{24}&$2$&\sg{53}&$2$&\sg{84}&$2$&\sg{113}&$2$&\sg{141}&$2$&\sg{181}&$3$&\sg{220}&$6$\\
\sg{26}&$2$&\sg{54}&$4$&\sg{85}&$2$&\sg{114}&$2$&\sg{142}&$4$&\sg{182}&$2$&\sg{222}&$2$\\
\sg{27}&$2$&\sg{55}&$2$&\sg{86}&$2$&\sg{116}&$2$&\sg{144}&$3$&\sg{184}&$2$&\sg{223}&$2$\\
\sg{28}&$2$&\sg{56}&$4$&\sg{88}&$2$&\sg{117}&$2$&\sg{145}&$3$&\sg{185}&$2$&\sg{224}&$2$\\
\sg{29}&$4$&\sg{57}&$4$&\sg{90}&$2$&\sg{118}&$2$&\sg{151}&$3$&\sg{186}&$2$&\sg{226}&$2$\\
\sg{30}&$2$&\sg{58}&$2$&\sg{91}&$4$&\sg{120}&$2$&\sg{152}&$3$&\sg{188}&$2$&\sg{227}&$2$\\
\sg{31}&$2$&\sg{59}&$2$&\sg{92}&$4$&\sg{122}&$2$&\sg{153}&$3$&\sg{190}&$2$&\sg{228}&$4$\\
\sg{32}&$2$&\sg{60}&$4$&\sg{93}&$2$&\sg{124}&$2$&\sg{154}&$3$&\sg{192}&$2$&\sg{230}&$8$\\
\sg{33}&$4$&\sg{61}&$4$&\sg{94}&$2$&\sg{125}&$2$&\sg{158}&$2$&\sg{193}&$2$& &\\
\sg{34}&$2$&\sg{62}&$4$&\sg{95}&$4$&\sg{126}&$2$&\sg{159}&$2$&\sg{194}&$2$& &\\
\sg{36}&$2$&\sg{63}&$2$&\sg{96}&$4$&\sg{127}&$2$&\sg{161}&$2$&\sg{198}&$4$& &\\
\sg{37}&$2$&\sg{64}&$2$&\sg{98}&$2$&\sg{128}&$2$&\sg{163}&$2$&\sg{199}&$4$& &\\
\end{tabular}
\end{center}
\end{table*}

\subsection{73 symmorphic space groups}
\sg{1-3, 5, 6 , 8, 10, 12, 16, 21-23, 25, 35, 38, 42, 44, 47, 65, 69, 71, 75, 79, 81, 82, 93, 87, 89, 97, 99, 197, 111, 115, 119, 121, 123, 139, 143, 146-150, 155-157, 160, 162, 164, 166, 168, 174, 175, 177, 183, 187, 189, 191, 195-197, 200, 202, 204, 207, 209, 211, 215-217, 221, 225, 229}

\subsection{62 centrosymmetric space groups}
\sg{2, 10-15, 47-74, 83-88, 123-142, 147, 148, 162-167, 175, 176, 191-194, 200-206, 221-230}

\clearpage
\section{Space groups allowing $M=2$~(Minimal connectivity of zero-frequency bands)}

\begin{table*}[h]
\begin{center}
\caption{List of 104 space groups that a band gap between the second and the third bands~($M=2$) of a time-reversal invariant \emph{dielectric} photonic crystal is not forbidden by the compatibility relations. They are listed according to their 16 supergroups.
$\cal{P}$ means centrosymmetric space groups~(with inversion) and $\cal{S}$ means symmorphic space groups.
\label{tabMeq2}}
\begin{tabular}{l|cc|c|c}\hline\hline
Key group&$\cal{P}$&$\cal{S}$ &$M$& $t$-subgroups\\\hline
\sg{39} ($Aem2$)&&& 2 & \sg{1, 5, 7, 8, 39}\\\hline
\sg{48} ($P\frac{2}{n}\frac{2}{n}\frac{2}{n}$)&$\cal{P}$&& 2 & \sg{1, 2, 3, 7, 13, 16, 34, 48}\\\hline
\sg{49} ($P\frac{2}{c}\frac{2}{c}\frac{2}{m}$)&$\cal{P}$&& 2 & \sg{1, 2, 3, 6, 7, 10, 13, 16, 27, 28, 49}\\\hline
\sg{50} ($P\frac{2}{b}\frac{2}{a}\frac{2}{n}$)&$\cal{P}$&& 2 &  \sg{1, 2, 3, 7, 13, 16, 30, 32, 50}\\ \hline
\sg{68} ($C\frac{2}{c}\frac{2}{c}\frac{2}{e}$)&$\cal{P}$&& 2 & \sg{1, 2, 3, 5, 7, 9, 13, 15, 21, 37, 41, 68}\\\hline
\sg{86} ($P4_{2}/n$)&$\cal{P}$&& 2 & \sg{1, 2, 3, 7, 13, 77, 81, 86}\\\hline
\sg{100} ($P4bm$)&&& 2 & \sg{1, 3, 7, 8, 32, 35, 75, 100}\\\hline
\sg{101} ($P4_2cm$)&&& 2 & \sg{1, 3, 7, 8, 27, 35, 77, 101}\\\hline
\sg{102} ($P4_2nm$)&&& 2 & \sg{1, 3, 7, 8, 34, 35, 77, 102}\\\hline
\sg{107} ($I4mm$)&&$\cal{S}$& 2 &\sg{1, 2, 5, 8, 42, 44, 79, 107}\\\hline
\sg{108} ($I4cm$)&&& 2 & \sg{1, 2, 5, 8, 9, 42, 45, 79, 108}\\\hline
\sg{131} ($P4_2/mmc$)&$\cal{P}$&& 2 & \sg{1, 2, 3, 5, 6, 9, 10, 15, 16, 21, 25, 37, 40, 47, 66, 77, 81, 84, 93,}\\&&&&\sg{105, 112, 115, 131} \\\hline
\sg{191} ($P6/mmm$)&$\cal{P}$&$\cal{S}$& 2 & \sg{1, 2, 3, 5, 6, 8, 10, 12, 35, 38, 143, 147, 149, 150, 156, 157, 162, 164}\\
&&&&\sg{168, 174, 175, 177, 183, 187, 189, 191}\\\hline
\sg{214} ($I4_{1}32$)&&& 2 & \sg{1, 5, 22, 24, 80, 98, 146, 155, 199, 214} \\\hline
\sg{221} ($Pm\bar{3}m$)&$\cal{P}$&$\cal{S}$& 2 & \sg{1, 2, 3, 5, 6, 8, 10, 12, 16, 21, 25, 35, 38, 47, 65, 75, 81, 83, 89, }\\
&&&&
\sg{99, 111, 115, 123, 146, 148, 155, 160, 166, 195, 200, 207, 215, 221}\\\hline
\sg{227} ($Fd\bar{3}m$)&$\cal{P}$&& 2 & \sg{1, 2, 5, 8, 9, 12, 15, 22, 24, 43, 44, 46, 70, 74, 80, 82, 88, 98, 109,}\\
&&&&
\sg{119, 122, 141, 146, 148, 155, 160, 166, 196, 203, 210, 216, 227} \\
\hline\hline
\multicolumn{5}{l}{}\\
\multicolumn{5}{l}{Space groups of $M=2$ allowed by compatibility relations:}\\
\multicolumn{5}{l}{
\sg{1}, \sg{2}, \sg{3}, \sg{5}, \sg{6}, \sg{7}, \sg{8}, \sg{9}, \sg{10}, \sg{12}, \sg{13}, \sg{15}, \sg{16}, \sg{21}, \sg{22}, \sg{24}, \sg{25}, \sg{27}, \sg{28},
\sg{30}, \sg{32}, \sg{34}, \sg{35}, \sg{37}, \sg{38}, \sg{\underline{39}}, \sg{40}, \sg{41}, \sg{42}, \sg{43}, \sg{44}, \sg{45},}\\
\multicolumn{5}{l}{
 \sg{46}, \sg{47}, \sg{\underline{48}}, \sg{\underline{49}}, \sg{\underline{50}}, \sg{65}, \sg{66}, \sg{\underline{68}}, \sg{70}, \sg{74}, \sg{75}, \sg{77}, \sg{79}, \sg{80}, \sg{81}, \sg{82}, \sg{83}, \sg{84}, \sg{\underline{86}}, \sg{88},
\sg{89}, \sg{93}, \sg{98}, \sg{99}, \sg{\underline{100}}, \sg{\underline{101}}, \sg{\underline{102}}, \sg{105},}\\
\multicolumn{5}{l}{
 \sg{\underline{107}}, \sg{\underline{108}}, \sg{109}, \sg{111}, \sg{112}, \sg{115}, \sg{119}, \sg{122}, \sg{123}, \sg{\underline{131}}, \sg{141}, \sg{143}, \sg{146}, \sg{147}, \sg{148}, \sg{149}, \sg{150}, \sg{155}, \sg{156},
\sg{157}, \sg{160}, \sg{162}, }\\
\multicolumn{5}{l}{
\sg{164}, \sg{166}, \sg{168}, \sg{174}, \sg{175}, \sg{177}, \sg{183}, \sg{187}, \sg{189}, \sg{\underline{191}},\sg{195}, \sg{196}, \sg{199}, \sg{200}, \sg{203}, \sg{207}, \sg{210}, \sg{\underline{214}}, \sg{215}, \sg{216}, \sg{\underline{221}}, \sg{\underline{227}}.}
\\
\end{tabular}
\end{center}
\end{table*}

\clearpage
\section{Space groups of $M>2$~(Minimal connectivity of zero-frequency bands)}
\begin{table}[h]
\begin{center}
\caption{List of 126 space groups of~($M>2$), in which a band gap between the second and third bands is forbidden by the compatibility relations in a time-reversal invariant \emph{dielectric} photonic crystals. There must be band crossings between the second and third bands along a high-symmetry momentum line.
We listed them according to their 22 $t$-subgroups.
$\cal{P}$ means centrosymmetric~(with inversion) and $\cal{S}$ means symmorphic~(without screw nor glide).
``$M\geqslant$'' represents the lower-bound of $M$.
\label{tabMneq2}}
\begin{tabular}{l|cc|c|c}\hline\hline
Key group &$\cal{P}$&$\cal{S}$& $M\geqslant$& $t$-supergroups\\\hline
\sg{4} ($P2_1$)&&& 4 &\sg{11,14,17-20,26,29,31,33,36,51-64}\\
&&&&\sg{76,78,90-92,94-96,113,114,127-130}\\
&&&&\sg{135-170,173,176,178,179,182}\\
&&&&\sg{185,186,193,194,198,205,212,213}\\ \hline
\sg{23} ($I222$)&&$\cal{S}$& 3
&\sg{71,72,97,121,139,140,197}\\
&&&&
\sg{204,209,211,217,225,226,229}\\ \hline
\sg{67} ($Cmme$)&$\cal{P}$&& 4 &\sg{125,129,134,138,224}\\ \hline
\sg{69} ($Fmmm$)&$\cal{P}$&$\cal{S}$& 3 &\sg{139,140,202,225,226,229}\\ \hline
\sg{73} ($Ibca$)&$\cal{P}$&& 4 &\sg{142,206,228,230}\\ \hline
\sg{85} ($P4/n$)&$\cal{P}$&& 4 &\sg{125,126,129,130,222}\\ \hline
\sg{87} ($I4/m$)&$\cal{P}$&$\cal{S}$& 3 &\sg{139,140,225,226,229}\\ \hline
\sg{103} ($P4cc$)&&& 4 &\sg{124,130}\\ \hline
\sg{104} ($P4nc$)&&& 4 &\sg{126,128,222}\\ \hline
\sg{106} ($P4_2bc$)&&& 4 &\sg{133,135}\\ \hline
\sg{110} ($I4_1cd$)&&& 4 &\sg{142,228,230}\\ \hline
\sg{116} ($P\bar{4}c2$)&&& 4 &\sg{124,130,132,138}\\ \hline
\sg{117} ($P\bar{4}b2$)&&& 4 &\sg{125,127,133,135}\\ \hline
\sg{118} ($P\bar{4}n2$)&&& 4 &\sg{126,128,134,136,222,224}\\ \hline
\sg{120} ($I\bar{4}c2$)&&& 4 &\sg{140,142,219,226,228,230}\\ \hline
\sg{144} ($P3_1$)&&& 3 &\sg{151,152,169,172,178,181}\\ \hline
\sg{145} ($P3_2$)&&& 3 &\sg{153,154,170,171,179,180}\\ \hline
\sg{158} ($P3c1$)&&& 4 &\sg{165,184,185,188,192,193}\\ \hline
\sg{159} ($P31c$)&&& 4 &\sg{163,184,186,190,192,194}\\ \hline
\sg{161} ($R3c$)&&& 4 &\sg{167,218-220,222,223,226,228,230}\\ \hline
\sg{201} ($Pn3$)&$\cal{P}$&& 4 &\sg{222,224}\\ \hline
\sg{208} ($P4_232$)&&& 4 &\sg{223,224}\\
\hline\hline
\multicolumn{5}{l}{}\\
\multicolumn{5}{l}{Space groups of $M>2$ :}\\
\multicolumn{5}{l}{
\sg{\underline{4}}, \sg{11}, \sg{14}, \sg{17}, \sg{18}, \sg{19}, \sg{20}, \sg{\underline{23}}, \sg{26}, \sg{29}, \sg{31}, \sg{33}, \sg{36}, \sg{51}, \sg{52}, \sg{53}, \sg{54}, \sg{55}, \sg{56},
\sg{57}, \sg{58}, \sg{59}, \sg{60}, \sg{61}, \sg{62}, \sg{63}, \sg{64}, \sg{\underline{67}}, \sg{\underline{69}}, \sg{71}, \sg{72},}\\
\multicolumn{5}{l}{
\sg{\underline{73}}, \sg{76}, \sg{78}, \sg{\underline{85}}, \sg{\underline{87}}, \sg{90}, \sg{91}, \sg{92}, \sg{94}, \sg{95}, \sg{96}, \sg{97}, \sg{\underline{103}}, \sg{\underline{104}}, \sg{\underline{106}}, \sg{\underline{110}}, \sg{113}, \sg{114}, \sg{\underline{116}}, \sg{\underline{117}}, \sg{\underline{118}},
\sg{\underline{120}}, \sg{121}, \sg{124}, \sg{125}, \sg{126},}\\
\multicolumn{5}{l}{
\sg{127}, \sg{128}, \sg{129}, \sg{130}, \sg{132}, \sg{133}, \sg{134}, \sg{135}, \sg{136},\sg{137}, \sg{138}, \sg{139}, \sg{140}, \sg{142}, \sg{\underline{144}}, \sg{\underline{145}}, \sg{151}, \sg{152}, \sg{153}, \sg{154}, \sg{\underline{158}}, \sg{\underline{159}},
\sg{\underline{161}},}\\
\multicolumn{5}{l}{
\sg{163}, \sg{165}, \sg{167}, \sg{169}, \sg{170}, \sg{171}, \sg{172}, \sg{173}, \sg{176}, \sg{178}, \sg{179}, \sg{180},\sg{181}, \sg{182}, \sg{184}, \sg{185}, \sg{186}, \sg{188}, \sg{190}, \sg{192}, \sg{193}, \sg{194}, \sg{197},}\\
\multicolumn{5}{l}{
\sg{198}, \sg{\underline{201}}, \sg{202}, \sg{204}, \sg{205}, \sg{206}, \sg{\underline{208}}, \sg{209}, \sg{211}, \sg{212}, \sg{213}, \sg{217}, \sg{218}, \sg{219}, \sg{220}, \sg{222}, \sg{223}, \sg{224}, \sg{225}, \sg{226}, \sg{228}, \sg{229}, \sg{230}.}
\\
\end{tabular}
\end{center}
\end{table}

\clearpage

\section{Compatibility relations.}
\label{app:CR}
Consider a crystal with a space group $\mathcal{G}$.  An element $g\in\mathcal{G}$ maps $\vec{r}=(x,y,z)$ to $p_g\vec{r}+\vec{t}_g$ where $p_g\in O(3)$ is a $3$ by $3$ orthogonal matrix.  For each $\vec{k}$ in the Brillouin zone, we define the little group $\mathcal{G}_{\vec{k}}=\{g\in\mathcal{G}\,|\,p_g\vec{k}=\vec{k} \text{ mod }\vec{G}\}$ that changes $\vec{k}$ only by a reciprocal lattice vector $\vec{G}=\sum_{\alpha=1}^3n_\alpha\vec{b}_\alpha$. Here, $n_\alpha$ are integers and $\vec{b}_\alpha$ are the reciprocal primitive vectors.

The wavefunctions at a high-symmetry point $K$ ($=\Gamma, X, Z,\ldots$) in the Brillouin zone belong to irreducible representations $U_i^K$ ($i=1,2,\ldots$) of $\mathcal{G}_K$.  The dimension of the irreducible representation $\text{dim}[U_i^K]$ generically indicates the order of the degeneracy at the point $K$, but the degeneracy might be enhanced due to the time-reversal symmetry.  The full list of irreducible representations for each space group $\mathcal{G}$ and each high-symmetry momentum $K$ is available in Ref.~\cite{Bradley}.  

Let us consider a set of $M$ bands separable from both higher and lower bands by full band gaps.  Suppose that an irreducible representation $U_i^K$ appears $n_i^K$ $(\geqslant0)$ times in these bands.  By definition $\sum_in_i^K \text{dim}[U_i^K]=M$.  The possible combination of integers $\{n_i^K\}_{i=1,2,\cdots}$ at two high-symmetry points $K=K_1, K_2$ are constrained by the symmetry along the line(s) connecting $K_1$ and $K_2$, as we will see below through several examples.  These constraints are called the ``compatibility relations" and one has to check them for every combination of high-symmetry momenta.  The compatibility relations in turn restrict the possible values of $M$ --- one cannot separate an arbitrary number of bands as there may not be any solution to the compatibility relations for a given $M$~\cite{usPRL}.

\section{The absence of the $M=2$ band gap for the key 22 space groups}
\label{app:nogo}

As we explained in the main text, we need to prove that a full band gap at $M=2$ is prohibited for the following 22 space groups:
\begin{quote}
\sg{4}, \sg{23}, \sg{67}, \sg{69}, \sg{73}, \sg{85}, \sg{87}, \sg{103}, \sg{104}, \sg{106}, \sg{110}, \sg{116}, \sg{117}, \sg{118}, \sg{120}, \sg{144}, \sg{145}, \sg{158}, \sg{159}, \sg{161}, \sg{201}, \sg{208}.
\end{quote}
In the main text, we presented the proof for \sg{4} and \sg{23}.  Also, the argument in Ref.~\cite{usPRL} that disproves a full band gap at $M=2$ can be applied to \sg{73}, \sg{106}, \sg{110}, \sg{144}, and \sg{145}, since the argument did not involve the singular point $\Gamma$.  In the following, we will discuss the remaining 15 space groups.

\subsection{\sg{103}, \sg{104}, \sg{158}, \sg{159}, and \sg{161}}
For these five space groups, the only line one should look at is the one connecting $\Gamma$ and $Z=(0,0,\frac{\pi}{c})$.  (For \sg{158} and \sg{159} belonging to the hexagonal lattice, the $Z$ point is called the $A$ point.)  There are several 1D representations and one 2D representation all the way along this line~\cite{Bradley}, and the two gapless photons belong to the 2D representation because of the rotation eigenvalue. At the $Z$ point, the 2D representation must appear twice to implement the time-reversal symmetry~\cite{Bradley}, but that requires in total 4 bands. Therefore, $M=2$ is prohibited.

\vspace{1\baselineskip}
To prove the absence of $M=2$ band gaps by contradiction for the remaining space groups, we will assume a full gap between the second and the third bands and then derive a contradiction based on the wrong assumption.  

\subsection{67}
The space group \sg{67} ($Cmme$)  belongs to the base-centered orthorhombic system with the primitive lattice vectors
\begin{eqnarray}
\vec{a}_1&=&\textstyle\frac{1}{2}(a,-b,0)\label{fcos1},\\
\vec{a}_2&=&\textstyle\frac{1}{2}(a,b,0),\\
\vec{a}_3&=&\textstyle(0,0,c).
\end{eqnarray}
The corresponding primitive reciprocal lattice vectors are
\begin{eqnarray}
\vec{b}_1&=&(\textstyle\frac{2\pi}{a},-\frac{2\pi}{b},0),\\
\vec{b}_2&=&(\textstyle\frac{2\pi}{a},\frac{2\pi}{b},0,\\
\vec{b}_3&=&(0,0,\textstyle\frac{2\pi}{c})\label{fcos2}. 
\end{eqnarray}
The space group is generated by a $\pi$-rotation $C_{2x}$, a screw $S_{2y}$, the inversion $I$,
\begin{eqnarray}
C_{2x}&:& (x,y,z)\mapsto(x,-y,-z),\\
S_{2y}&:& (x,y,z)\mapsto\textstyle(-x,y+\frac{b}{2},-z),\\
I&:& (x,y,z)\mapsto -(x,y,z),
\end{eqnarray}
and the lattice translations by $\vec{a}_1$, $\vec{a}_2$, and $\vec{a}_3$.  The $\Gamma=(0,0,0)$, $Y=(0,\frac{2\pi}{b},0)$, $Z=(0,0,\frac{\pi}{c})$, and $T=(0,\frac{2\pi}{b},\frac{\pi}{c})$ points have all of these symmetries and the representations are all one-dimensional at these points~\cite{Bradley}.  We examine the several lines among them. 

There are two lines connecting $\Gamma$ and $Y$, $(0,k_y,0)$ ($k_y\in[0,\frac{2\pi}{b}]$) symmetric under $S_{2y}$ and $C_{2x}I$ and $(k_x,0,0)$ ($k_x\in[0,\frac{2\pi}{a}]$) symmetric under $C_{2x}$ and $S_{2y}I$   (recall that $(\frac{2\pi}{a},0,0)=Y+\vec{b}_1$ is equivalent with $Y$). These lines states that one of the two gapless photons has $(C_{2x},S_{2y},I)=(-1,+1,+1)$ and the other has $(C_{2x},S_{2y},I)=(-1,+1,-1)$ at $Y$.  Both of these modes have the $-1$ eigenvalue of $S_{2y}C_{2x}$, while one of the two modes have the eigenvalue $+1$ and the other has $-1$ eigenvalue of $C_{2x}I$.

Now consider the line $(0,\frac{2\pi}{b},k_z)$  ($k_z\in[0,\frac{\pi}{c}]$) from $Y$ to $T$, symmetric under $S_{2y}C_{2x}$ and $C_{2x}I$. The number of $\pm1$ eigenvalues of these symmetries must be conserved along this line. Therefore, one of the two modes has $(C_{2x},S_{2y},I)=(\xi_1,-\xi_1,\xi_1)$ ($\xi_1^2=1$), and the other has  $(C_{2x},S_{2y},I)=(\xi_2,-\xi_2,-\xi_2)$ ($\xi_2^2=1$) at $T$.

Next, we consider the line $(0,0,k_z)$ ($k_z\in[0,\frac{\pi}{c}]$) from $\Gamma$ to $Z$, symmetric under $S_{2y}C_{2x}$ and $C_{2x}I$. From the same reason, one of the two modes have $(C_{2x},S_{2y},I)=(\xi_3,-\xi_3,\xi_3)$ ($\xi_3^2=1$), and the other has  $(C_{2x},S_{2y},I)=(\xi_4,-\xi_4,-\xi_4)$ ($\xi_2^2=1$) at $Z$.

Finally, we look at two lines connecting $Z$ and $T$, $(0,k_y,\frac{\pi}{c})$ ($k_y\in[0,\frac{\pi}{b}]$) symmetric under $S_{2y}$ and $C_{2x}I$ and $(k_x,0,\frac{2\pi}{c})$ ($k_x\in[0,\frac{2\pi}{a}]$) symmetric under $C_{2x}$ and $S_{2y}I$. By the conservation of the eigenvalues, we have
\begin{eqnarray}
S_{2y}&:&-\xi_1-\xi_2=-(-\xi_3-\xi_4),\\
C_{2x}I&:&0=0,\\
S_{2y}C_{2x}I&:&-\xi_1+\xi_2=-(-\xi_3+\xi_4)
\end{eqnarray}
and
\begin{eqnarray}
C_{2x}&:&\xi_1+\xi_2=\xi_3+\xi_4,\\
S_{2y}I&:&0=0,\\
S_{2y}C_{2x}I&:&-\xi_1+\xi_2=-\xi_3+\xi_4.
\end{eqnarray}
The unique solution to these simultaneous equations are $\xi_i=0$, which violates $\xi_i^2=+1$. This is a contradiction.

\subsection{69}
The space group \sg{69} ($Fmmm$)  belongs to the face-centered orthorhombic system with the primitive lattice vectors
\begin{eqnarray}
\vec{a}_1&=&\textstyle\frac{1}{2}(0,b,c)\label{fcos1},\\
\vec{a}_2&=&\textstyle\frac{1}{2}(a,0,c),\\
\vec{a}_3&=&\textstyle\frac{1}{2}(a,b,0).
\end{eqnarray}
The corresponding primitive reciprocal lattice vectors are
\begin{eqnarray}
\vec{b}_1&=&(-\textstyle\frac{2\pi}{a},\frac{2\pi}{b},\frac{2\pi}{c}),\\
\vec{b}_2&=&(\textstyle\frac{2\pi}{a},-\frac{2\pi}{b},\frac{2\pi}{c}),\\
\vec{b}_3&=&(\textstyle\frac{2\pi}{a},\frac{2\pi}{b},-\frac{2\pi}{c})\label{fcos2}.
\end{eqnarray}
The space group is generated by $\pi$-rotations $C_{2x}$, $C_{2y}$ about $x$, $y$ axes, the inversion $I$
\begin{eqnarray}
C_{2x}&:& (x,y,z)\mapsto(x,-y,-z),\\
C_{2y}&:& (x,y,z)\mapsto(-x,y,-z),\\
I&:& (x,y,z)\mapsto -(x,y,z),
\end{eqnarray}
and the lattice translations by $\vec{a}_1$, $\vec{a}_2$, and $\vec{a}_3$.  The $\Gamma=(0,0,0)$, $X=(\frac{2\pi}{a},0,0)$, $Y=(0,\frac{2\pi}{b},0)$, and $Z=(0,0,\frac{2\pi}{c})$ points have all of these symmetries.  We examine the several lines between these points. 

There are two lines connecting $\Gamma=(0,0,0)$ and $X=(\frac{2\pi}{a},0,0)$.  The line $(k_x,0,0)$, symmetric under $C_{2x}$, $IC_{2y}$, and $IC_{2z}$ ($C_{2z}\equiv C_{2x}C_{2y}$), demands that the two linear gapless modes have (i) two $-1$ eigenvalues of $C_{2x}$ and (ii) one $+1$ and one $-1$ eigenvalues of $IC_{2y}$ and $IC_{2z}$ at $X$.  Also, the line $(0,k,k)$ between $\Gamma$ and $X+\vec{b}_1=(0,\frac{2\pi}{b},\frac{2\pi}{c})$ requires that one of the two modes have one $+1$ and the other has $-1$ eigenvalue of $IC_{2x}$ at $X$.  This, in turn, means that the eigenvalue of $C_{2y}$ of the two modes are the same.  Therefore, both of the two modes have the same rotation eigenvalue $(-1,\zeta_y,\zeta_z)$ with $\zeta_z=-\zeta_y=\pm1$ at $X$.  One can derive similar conditions for the $Y$ and $Z$ points in the same way; i.e., $(\zeta_x',-1,\zeta_z')$ at $Y$ and $(\zeta_x'',\zeta_y'',-1)$ at $Z$

There is also a line $\vec{k}=(\frac{2\pi}{a},k_y,0)$ connecting $X=(\frac{2\pi}{a},0,0)$ and $Z+\vec{b}_3=(\frac{2\pi}{a},\frac{2\pi}{b},0)$, which means that the eigenvalues of $C_{2y}$ at $X$ and $Z$ are identical, $\zeta_y''=\zeta_y$.  Similarly, $\zeta_x''=\zeta_x'$ and $\zeta_z'=\zeta_x$.  All in all, both of the two modes have the following eigenvalues of the three $\pi$-rotations,
\begin{eqnarray}
X&&: (-1,\zeta_y,\zeta_z),\\
Y&&: (\zeta_x,-1,\zeta_z),\\
Z&&: (\zeta_x,\zeta_y,-1).
\end{eqnarray}
Here, $\zeta_\alpha=\pm1$ must satisfy $\zeta_y\zeta_z=-1$,  $\zeta_z\zeta_x=-1$, and $\zeta_x\zeta_y=-1$, but they cannot hold simultaneously. This is a contradiction.

\subsection{\sg{85}}
The space group \sg{85} ($P4/n$) belongs to the primitive tetragonal lattice system with the primitive lattice vectors and the primitive reciprocal lattice vectors
\begin{eqnarray}
\vec{a}_1&=&(a,0,0),\label{ptls1}\\
\vec{a}_2&=&(0,a,0),\\
\vec{a}_3&=&(0,0,c).
\end{eqnarray}
\begin{eqnarray}
\vec{b}_1&=&\textstyle(\frac{2\pi}{a},0,0),\\
\vec{b}_2&=&\textstyle(0,\frac{2\pi}{a},0),\\
\vec{b}_3&=&\textstyle(0,0,\frac{2\pi}{c}).\label{ptls2}
\end{eqnarray}
The group is generated by 
\begin{eqnarray}
C_{4z}&:& (x,y,z)\mapsto(-y+\textstyle\frac{a}{2},x,z),\\
I&:& (x,y,z)\mapsto -(x,y,z),
\end{eqnarray}
and the lattice translations.

Along the line $(0,0,k_z)$ ($k_z\in[0,\frac{\pi}{c}]$) connecting $\Gamma$ and $Z=(0,0,\frac{\pi}{c})$, each band can be labeled by the eigenvalue of $C_{4z}$.  At $Z=(0,0,\frac{\pi}{c})$, one of the two linear gapless modes must have the eigenvalue $(C_{4z},I)=(+ i,\xi)$ ($\xi=\pm1$) and the other must have $(C_{4z},I)=(-i,\xi)$ due to the time-reversal symmetry.  Both of the two modes have the eigenvalue $(\pm i)^2\xi=-\xi$ of the glide $G_z\equiv C_{4z}^2I: (x,y,z)\mapsto(x+\frac{a}{2},y+\frac{a}{2},-z)$.

Next, we look at the $R=(\frac{\pi}{a},0,\frac{\pi}{c})$ point, where $G_z$ and $I$ generates the $\mathbb{Z}_2\times\mathbb{Z}_2$ symmetry.  Note that $G_z$ and $I$ do not commute at $R$; they satisfy $G_zI=T_xT_yIG_z=-IG_z$ where $T_x$ and $T_y$ are translations in $x,y$ by $a$ and hence take the value $T_x=+1$ and $T_y=-1$ at $R$.  Their 2D representation is given by the Pauli matrix, which are traceless except for the identity.

Finally, we consider the line $(k_x,0,\frac{\pi}{c})$ ($k_x\in[0,\frac{\pi}{a}]$) connecting $Z$ and $R$. The line is symmetric under $G_z$.  The eigenvalue $-\xi$ of $G_z$ at $Z$ becomes $+\xi$ at $R$ due to the nonsymmorphic nature of $G_z$. However, this is still in contradiction since the 2D representation at $R$ is traceless.  Therefore, there must be at least two more bands that have the eigenvalues $(C_{4z},I)=(\pm i,-\xi)$ at $\Gamma$. In total four bands must cross with each other along $\Gamma$-$Z$-$R$.

\subsection{\sg{87}}
The space group \sg{87} ($I4/m$) belongs to the body-centered tetragonal lattice system with the primitive lattice vectors and the reciprocal vectors
\begin{eqnarray}
\vec{a}_1&=&\textstyle\frac{1}{2}(-a,a,c),\label{btls1}\\
\vec{a}_2&=&\textstyle\frac{1}{2}(a,-a,c),\\
\vec{a}_3&=&\textstyle\frac{1}{2}(a,a,-c).
\end{eqnarray}
\begin{eqnarray}
\vec{b}_1&=&\textstyle(0,\frac{2\pi}{a},\frac{2\pi}{c}),\\
\vec{b}_2&=&\textstyle(\frac{2\pi}{a},0,\frac{2\pi}{c}),\\
\vec{b}_3&=&\textstyle(\frac{2\pi}{a},\frac{2\pi}{a},0).\label{btls2}
\end{eqnarray}
The group is generated by 
\begin{eqnarray}
C_{4z}&:& (x,y,z)\mapsto(-y,x,z),\\
I&:& (x,y,z)\mapsto -(x,y,z),
\end{eqnarray}
and the lattice translations.

We have to look at two lines connecting $\Gamma=(0,0,0)$ and $Z=(0,0,\frac{2\pi}{c})$.  The first line is $(0,0,k_z)$ ($k_z\in [0,\frac{2\pi}{c}]$), which requires that, just as in the case for \sg{85}, one of the two linear gapless modes has the eigenvalue $(C_{4z},I)=(+ i,\xi)$ and the other has $(C_{4z},I)=(-i,\xi)$ ($\xi=\pm1$) due to the time-reversal symmetry at $Z$.  Hence, their eigenvalue of the mirror $M_z\equiv IC_{4z}^2$ at $Z$ is $\xi (\pm i)^2=-\xi$.

The second line is $(k_x,0,0)$ ($k_x\in [0,\frac{2\pi}{a}]$) (recall  that $(\frac{2\pi}{a},0,0)=Z-\vec{b}_1+\vec{b}_3$ is equivalent with $Z$), which is symmetric under the mirror $M_z$.  Therefore, the two modes must have one $+1$ and one $-1$ eigenvalue of $M_z$. However, this contradicts with the fact that the two modes have the same eigenvalue $-\xi$ of $M_z$.

\subsection{\sg{116}}
The space group \sg{116} ($P\bar{4}c2$) belongs to the primitive tetragonal lattice system [Eqs.~\eqref{ptls1}-\eqref{ptls2}]. The group is generated by
\begin{eqnarray}
\bar{C}_{4z}&:& (x,y,z)\mapsto(y,-x,-z),\\
G_y&:& (x,y,z)\mapsto\textstyle(x,-y,z+\frac{c}{2}),
\end{eqnarray}
and the lattice translations.

Consider the path from $\Gamma$ to  $A=\textstyle(\frac{\pi}{a},\frac{\pi}{a},\frac{\pi}{c})$ via $M=(\frac{\pi}{a},\frac{\pi}{a},0)$.  The line $(k,k,0)$ ($k\in[0,\frac{\pi}{a}]$) between $\Gamma$ and $M$ requires that the two gapless photons have the $-1$ eigenvalue of the $\pi$-rotation $G_y\bar{C}_{4z}: (x,y,z)\mapsto(y,x,\frac{c}{2}-z)$ at $M$.  There are four 1D representations and one 2D representation at $M$, and representations consistent with this rotation eigenvalue are the two 1D representations with $(\bar{C}_{4z}, G_y)=(-1,1)$ and $(1,-1)$~\cite{Bradley}, for which $\bar{C}_{4z}^2=+1$. Then the line $(\frac{\pi}{a},\frac{\pi}{a},k_z)$ ($k_z\in[0,\frac{\pi}{c}]$) between $M$ and $A$ indicates that the two modes belong to the 2D representation at $A$, since that is the only representation with $\bar{C}_{4z}^2=+1$ at $A$~\cite{Bradley}. This 2D representation is traceless except for the identity and $\bar{C}_{4z}^2$~\cite{Bradley}.

There is another route going to $A$. 
The line $(0,0,k_z)$ ($k_z\in[0,\frac{\pi}{c}]$) between $\Gamma$ and $Z=(0,0,\frac{\pi}{c})$ demands that one of the two modes has the 1D representation with $(\bar{C}_{4z},G_y)=(+ i,i \xi)$ ($\xi=\pm 1$) and the other has $(\bar{C}_{4z},G_y)=(-i,-i \xi)$ at $Z$. Hence, the two modes have the same eigenvalue of $G_y\bar{C}_{4z}=\xi$. Since the line $(k,k,\frac{\pi}{c})$ connecting $Z$ and $A$ is symmetric under $G_y\bar{C}_{4z}$, this contradicts with the traceless nature of the 2D representation.

The argument for the space group \sg{118} ($P\bar{4}n2$) is more or less identical.   The group also belongs to the primitive tetragonal lattice system [Eqs.~\eqref{ptls1}-\eqref{ptls2}] and is generated by
\begin{eqnarray}
\bar{C}_{4z}&:& (x,y,z)\mapsto(y,-x,-z),\\
G_y&:& (x,y,z)\mapsto\textstyle(\frac{a}{2}+x,\frac{a}{2}-y,\frac{a}{2}+z),
\end{eqnarray}
The path from $\Gamma$ to $A=\textstyle(\frac{\pi}{a},\frac{\pi}{a},\frac{\pi}{c})$ via $M=(\frac{\pi}{a},\frac{\pi}{a},0)$ demands that the two linear gapless modes have two 1D representations with $(\bar{C}_{4z}, G_y)=(-i,i)$ and $(i,-i)$ at $M$ and that they belong to the 2D representation at $A$ because of the $\bar{C}_{4z}^2=-1$ eigenvalue.  Then the second path from $\Gamma$ to $A$ via $Z$ cannot satisfy the traceless nature of the 2D representation at $A$.

\subsection{\sg{117}}
The space group \sg{117} ($P\bar{4}b2$) belongs to the primitive tetragonal lattice system [Eqs.~\eqref{ptls1}-\eqref{ptls2}]. The group is generated by
\begin{eqnarray}
\bar{C}_{4z}&:& (x,y,z)\mapsto(y,-x,-z),\\
G_y&:& (x,y,z)\mapsto\textstyle(\frac{a}{2}+x,\frac{a}{2}-y,z),
\end{eqnarray}
and the lattice translations.

Consider the path from $\Gamma$ to $A=\textstyle(\frac{\pi}{a},\frac{\pi}{a},\frac{\pi}{c})$ via $Z=\textstyle(0,0,\frac{\pi}{c})$.
The line $(0,0,k_z)$ ($k_z\in[0,\frac{\pi}{c}]$) between $\Gamma$ and $Z$ demands that both of the two gapless photons have the $-1$ eigenvalue of the $\pi$-rotation $\bar{C}_{4z}^2$.  Thus the two modes belong to the 2D representation at $Z$, since that is the only representation with $\bar{C}_{4z}^2=-1$~\cite{Bradley}. This 2D representation is traceless except for the identity and $\bar{C}_{4z}^2$~\cite{Bradley}.  Then the line $(k,k,\frac{\pi}{c})$ connecting $Z$ and $A$, symmetric under a screw $G_y\bar{C}_{4z}: (x,y,z)\mapsto(y+\frac{a}{2},x+\frac{a}{2},-z)$, tells us that the screw eigenvalues $+1$ and $-1$ come in pair at $A$.

There is another route going to $A$ from $\Gamma$. 
The line $(k,k,0)$ ($k\in[0,\frac{\pi}{a}]$) between $\Gamma$ and $M=(\frac{\pi}{a},\frac{\pi}{a},0)$ requires that the two modes have the $+1$ eigenvalue of the screw $G_y\bar{C}_{4z}$ at $M$.  The $M$ point have four 1D representations and one 2D representation. Among them, those consistent with this requirement of the screw eigenvalue are the two 1D representations $(\bar{C}_{4z},G_y)=(i,-i)$ and $(\bar{C}_{4z},G_y)=(-i,i)$, which come in pair due to the time-reversal symmetry. Lastly, the line $(\frac{\pi}{a},\frac{\pi}{a},k_z)$ ($k_z\in[0,\frac{\pi}{c}]$)  is symmetric under $\bar{C}_{4z}^2$ and $G_y$, and the eigenvalues of these symmetries ($\bar{C}_{4z}^2=(\pm i)^2 = -1$ and $G_z=\pm i$) are preserved along this line. 

There are four 1D representations and one 2D representation at $A$, and the 2D representation is inconsistent with the negative eigenvalue of $\bar{C}_{4z}^2$~\cite{Bradley}.  The four 1D representations at $A$ are labeled by $(\bar{C}_{4z},G_y)=(i \xi_1,i \xi_2)$ with $\xi_1=\pm 1$ and $\xi_2=\pm 1$. Due to the time-reversal symmetry at $A$, the representation $(i \xi_1,i \xi_2)$ must come with $(-i \xi_1,-i \xi_2)$. But then they have the same eigenvalue $-\xi_1\xi_2$ of $G_y\bar{C}_{4z}$. This is in contradiction with our conclusion of the first path that the screw eigenvalues $+1$ and $-1$ come in pair at $A$.

\subsection{\sg{120}}
The space group \sg{120} ($I\bar{4}c2$) has the same symmetries $\bar{C}_{4z}$, $G_y$ as \sg{116} but belongs to the body-centered tetragonal lattice system  [Eqs.~\eqref{btls1}-\eqref{btls2}].

The $P=(\frac{\pi}{a},\frac{\pi}{a},\frac{\pi}{c})$ point is not time-reversal invariant, but still the time-reversal symmetry $\mathcal{T}$ has a nontrivial consequence. The $P$ point is invariant under the $\bar{C}_{4z}$ symmetry and there are four 1D representations labeled by $\bar{C}_{4z}=\lambda$ ($\lambda^4=+1$). Consider the combined symmetry $\mathcal{T}'=\mathcal{T}G_y$ that preserves the $P$ point modulo a reciprocal lattice vector. $\mathcal{T}'$ is an anti-unitary symmetry that squares into $(\mathcal{T}')^2=G_y^2=T_z=-1$ at $P$.  Therefore, the band structure always exhibits two fold degeneracy at $P$.

Moreover, if $|\lambda\rangle$ has the eigenvalue $\lambda$ of $\bar{C}_{4z}$ (i.e., $\bar{C}_{4z}|\lambda\rangle=\lambda |\lambda\rangle$), then $\mathcal{T}'|\lambda\rangle$ has the eigenvalue $-(\lambda^*)^3$ of $\bar{C}_{4z}$. Indeed, using $\bar{C}_{4z}G_y=T_z^{-1}G_y\bar{C}_{4z}^3$, we get
\begin{equation}
\bar{C}_{4z}(\mathcal{T}'|\lambda\rangle)=\mathcal{T}(T_z^{-1}G_y\bar{C}_{4z}^3|\lambda\rangle)=-(\lambda^*)^3(\mathcal{T}'|\lambda\rangle).
\end{equation}
Therefore, the representation $\lambda$ always comes with $-(\lambda^*)^3$. Namely, not only $+i$ and $-i$ are paired, but $+1$ and $-1$ are also paired under $\mathcal{T}'$. As a result, the two bands have the same eigenvalue of $\bar{C}_{4z}^2=\pm1$, i.e., $\bar{C}_{4z}^2=-1$ for the $\pm i$ pair, and $\bar{C}_{4z}^2=+1$ for the $\pm 1$ pair. 

On the other hand, $X=(\frac{\pi}{a},\frac{\pi}{a},0)$ is invariant under $\bar{C}_{4z}^2$ and $G_z\bar{C}_{4z}$, which commute at $X$.
There are four 1D representations $(\bar{C}_{4z}^2, G_z\bar{C}_{4z})=(\xi_1,\xi_2)$ ($\xi_1^2=\xi_2^2=1$).  Lines connecting $\Gamma$, $X$, and $Z=(0,0,\frac{2\pi}{c})$ require that one of the two gapless photons has the representation $(\bar{C}_{4z}^2, G_z\bar{C}_{4z})=(+1,-1)$ and the other one has $(-1,-1)$ at $X$.

Finally, the line $(\frac{\pi}{a},\frac{\pi}{a},k_z)$ connecting $X$ and $P$ is invariant under $\bar{C}_{4z}^2$.  Hence, the number of eigenvalues $\pm1$ of $\bar{C}_{4z}^2$ must be preserved along this line. However, there are one $+1$ and one $-1$ eigenvalues at $X$, but there are two $+1$ or $-1$ eigenvalues at $P$, provided the $M=2$ gap. This is a contradiction.

\subsection{\sg{201}}
The space group \sg{201} ($Pn\bar{3}$) belongs to the primitive cubic lattice system [Eqs.~\eqref{ptls1}-\eqref{ptls2} with $c=a$].
The group is generated by
\begin{eqnarray}
C_{2z}&:& (x,y,z)\mapsto\textstyle(-x+\frac{a}{2},-y+\frac{a}{2},z),\\
C_{3}&:& (x,y,z)\mapsto(z,x,y),\\
I&:& (x,y,z)\mapsto -(x,y,z),
\end{eqnarray}

Let us start with the line connecting $\Gamma$ and $R=(\frac{\pi}{a},\frac{\pi}{a},\frac{\pi}{a})$ invariant under $C_3$.
There are 1D and 3D representations at $R$~\cite{Bradley}, but if we assume $M=2$ gap, the 3D representations are irrelevant. The diagonal line suggests that one of the two gapless photons has the 1D representation with $(C_{2z},C_{3},I)=(1,\omega,\xi)$ ($\xi=\pm 1$) and the other has $(C_{2z},C_{3},I)=(1,\omega^2,\xi)$, where $\omega^3=1$.  

Next, the line connecting $R=(\frac{\pi}{a},\frac{\pi}{a},\frac{\pi}{a})$ and $M=(\frac{\pi}{a},0,\frac{\pi}{a})$ symmetric under $C_{2z}I$. Along this line, the number of eigenvalues of $C_{2z}I$ must be conserved. At $R$, both of the two modes have $C_{2z}I=\xi$. However, at $M$, there are only two 2D representations, both of which are traceless for $C_{2z}I$~\cite{Bradley}. Hence, there must be at least four bands connecting with each other along the line $\Gamma$-$R$-$M$.

\subsection{\sg{208}}
The space group \sg{208} ($P4_232$) belongs to the primitive cubic lattice system [Eqs.~\eqref{ptls1}-\eqref{ptls2} with $c=a$].
The group is generated by
\begin{eqnarray}
S_{4y}&:& (x,y,z)\mapsto\textstyle(z+\frac{a}{2},y+\frac{a}{2},-x+\frac{a}{2}),\\
C_{3}&:& (x,y,z)\mapsto(z,x,y),\\
I&:& (x,y,z)\mapsto -(x,y,z).
\end{eqnarray}

There are two 1D, one 2D, and two 3D representations at $R=(\frac{\pi}{a},\frac{\pi}{a},\frac{\pi}{a})$~\cite{Bradley}. The line $(k,k,k)$ connecting $\Gamma$ and $R$, invariant under  $C_{3}$, demands that the two linear gapless modes belong to the 2D representation, in which $S_{4y}^2$ is represented by identity~\cite{Bradley}.

There are four 1D and one 2D representations at $M=(\frac{\pi}{a},0,\frac{\pi}{a})$~\cite{Bradley}. The line $(k,0,k)$ connecting $\Gamma$ and $M$, invariant under $C_{3}^2S_{4y}C_{3}^2$, indicates that the two linear gapless modes belong to 1D representations, in which $S_{4y}^2=+1$~\cite{Bradley}.

The line $(\frac{\pi}{a},k_y,\frac{\pi}{a})$ connecting $M$ and $R$ is invariant under the screw $S_{4y}^2$. Since  $S_{4y}^4=T_y^2=e^{i2k_ya}$, the eigenvalues of $S_{4y}^2$ have the momentum dependence of $e^{ik_ya}$. Hence, the eigenvalues flip sign when moving from $M$ to $R$. However, both of the two modes have the eigenvalue $+1$ of $S_{4y}^2$ at $M$ and $R$. This is a contradiction.

\section{The possibility of the $M=2$ band gap for the 16 key space groups}

\subsection{\sg{131} and \sg{227}}
This is proven by examples.

\subsection{\sg{39}, \sg{100}, \sg{101}, \sg{102}, \sg{107}, and \sg{108}}
We prove this by showing that there exists a tight-binding model that has the same symmetry eigenvalues as required for photonic band structures. These tight-binding models produce bands that are gapped in the entire Brillouin.

As the simplest example, let us look at \sg{39} ($Aem2$) generated by
\begin{eqnarray}
G_{2x}&:& (x,y,z)\mapsto\textstyle(-x,y+\frac{b}{2},z),\\
C_{2z}&:& (x,y,z)\mapsto(-x,-y,z),
\end{eqnarray}
and lattice translations 
\begin{eqnarray}
\vec{a}_1&=&\textstyle(a,0,0),\\
\vec{a}_2&=&\textstyle\frac{1}{2}(0,b,-c),\\
\vec{a}_3&=&\textstyle\frac{1}{2}(0,b,c).
\end{eqnarray}
There are several lattice structures consistent with this symmetry. Here we look at the one with a site at $\vec{x}_1=(0,0,0)$ and another site at $\vec{x}_2=(0,\frac{b}{2},0)$ within the unit cell spanned by $\vec{a}_1$, $\vec{a}_2$, and $\vec{a}_3$. We put one $p_y$ orbital on each site and consider the tight-binding model of these two orbitals per unit cell.

Let us ask how these orbitals transform under the symmetry operations in order to determine the representation of the band structure of the tight-binding model.  In particular, we are interested in the representation at the $\Gamma$ point, where additional constraints are imposed for photonic crystals.
 
Under the $C_{2z}$ symmetry, the $p_y$ orbital on the site $\vec{x}_1$ just flips sign.  The $p_y$ orbital on $\vec{x}_2$ also flips sign and moves to $\vec{x}_2-\vec{a}_2-\vec{a}_3$. But since we are interested in the symmetry representation at the $\Gamma$ point at which translations are all represented trivially, we can just neglect this position shift. Therefore $C_{2z}$ is represented by $-\sigma_0$ ($\sigma_0$ is the identity matrix).  

Under the $G_{2x}$ symmetry, the $p_y$ orbital on $\vec{x}_1$ and the one of $\vec{x}_2$ interchange with each other.  Hence, the representation of $G_{2x}=\sigma_1$ ($\sigma_{1,2,3}$ are the Pauli matrices).

The line $(k_x,0,0)$ is symmetric under the mirror $G_{2x}C_{2z}$ represented by $\sigma_1(-\sigma_0)=-\sigma_1$.  As required for photonic crystals, these two bands have one $+1$ and one $-1$ eigenvalue of the mirror. The line $(0,0,k_z)$ is symmetric under $C_{2z}=-\sigma_0$. Again, as required, both of the two bands have the $-1$ eigenvalue of $C_{2z}$. One can verify the line $(0,k_y,0)$ too. Hence, the band structure of this tight-binding model fulfills all the requirements of the symmetry eigenvalues imposed for phonic crystals.

More generally, lattices consistent with the assumed space group symmetry are classified by Wyckoff positions. In the above discussion of \sg{39}, we put a $p$ orbital on each site of the Wyckoff position labeled $a$ in~\cite{ITC}. Similarly, the tight-binding mode built from one $p$ orbital of each lattice site of the Wyckoff position in Table~\ref{AI} generates a band structure that satisfies all symmetry requirements for photonic crystals.

\begin{table}
\begin{center}
\caption{Wyckoff positions used in the proof.\label{AI}}
\begin{tabular}{c c}\hline\hline
Space group No.&Wyckoff position\\ \hline
 \sg{39} & $a$ or $b$\\
 \sg{100} & $b$\\
 \sg{101} & $a$ or $b$\\
 \sg{102} & $a$\\
 \sg{107} & $b$\\
 \sg{108} & $b$\\\hline\hline
\end{tabular}
\end{center}
\end{table}


\subsection{\sg{48}, \sg{49}, \sg{50}, \sg{68}, \sg{86}, \sg{191}, \sg{214}, and \sg{221}}
\label{app:solcomp}
Here we present the solution of the compatibility relations consistent with a $M=2$ photonic gap. We list the trace of the irreducible representations $\text{tr}[U_i^K(g)]$. The symmetry element $g$ is arranged in the order of Ref.~\cite{ITC}.  In these tables, $\bar{x}$ means $-x$ and ``$*$" means that the momentum $K$ is not symmetric under the corresponding symmetry operation.

For example, let us look at Table~\ref{68}. According to Ref.~\cite{ITC}, the group has eight symmetry elements in addition to the translations.  The momentum $K=S=(\frac{\pi}{a},\frac{\pi}{b},0)$ is invariant under the following symmetry operations:
\begin{eqnarray}
(1)&:& (x,y,z)\rightarrow\textstyle(x,y,z),\\
(2)&:& (x,y,z)\rightarrow\textstyle(-x+\frac{a}{2},-y+\frac{b}{2},z),\\
(5)&:& (x,y,z)\rightarrow\textstyle(-x,-y+\frac{b}{2},\bar{z}+\frac{c}{2}),\\
(6)&:& (x,y,z)\rightarrow\textstyle(x+\frac{a}{2},y,-z+\frac{c}{2}),
\end{eqnarray}
where the numbers in parenthesis are assigned in Ref.~\cite{ITC}.  
At this momentum, the gapless photons belong to the 2D representation which is traceless except for the identity operation (1). Hence, in the entry of $K=S$ in Table~\ref{68}, we have $2$, $0$, $0$, and $0$ for the symmetry elements (1), (2), (5), and (6), respectively, and ``$*$" for (3), (4), (7), and (8).

\begin{table}[h!]
\begin{center}
\caption{Solutions of the compatibility relations for \sg{48} ($Pnnn$).  $\xi_1^2=\xi_2^2=\xi_3^2=1$, $\xi_1\xi_2\xi_3=1$.\label{48}}
\begin{tabular}{c|c}\hline\hline
High-sym. momentum&$\text{tr}[U(g)]$ in the order of Ref.~\cite{ITC}\\ \hline
$Y=(0,\frac{\pi}{b},0)$&$(2,0,\bar{2},0,0,0,0,0)$\\ \hline
$X=(\frac{\pi}{a},0,0)$&$(2,0,0,\bar{2},0,0,0,0)$\\\hline
$Z=(0,0,\frac{\pi}{c})$&$(2,\bar{2},0,0,0,0,0,0)$\\\hline
$U=(\frac{\pi}{a},0,\frac{\pi}{c})$&$(2,0,2\xi_2,0,0,0,0,0)$\\\hline
$T=(0,\frac{\pi}{b},\frac{\pi}{c})$&$(2,0,0,2\xi_1,0,0,0,0)$\\\hline
$S=(\frac{\pi}{a},\frac{\pi}{b},0)$&$(2,2\xi_3,0,0,0,0,0,0)$\\\hline
$R=(\frac{\pi}{a},\frac{\pi}{b},\frac{\pi}{c})$&$(1,\xi_3,\xi_2,\xi_1,1,\xi_3,\xi_2,\xi_1)$\\
&$(1,\xi_3,\xi_2,\xi_1,\bar{1},\bar{\xi}_3,\bar{\xi}_2,\bar{\xi}_1)$\\ \hline \hline
\end{tabular}
\end{center}
\end{table}

\begin{table}[h!]
\begin{center}
\caption{Solutions of the compatibility relations for \sg{49}  ($Pccm$).  $\xi_1^2=\xi_2^2=\xi_3^2=1$, $\xi_1\xi_2\xi_3=1$.\label{49}}
\begin{tabular}{c|c}\hline\hline
High-sym. momentum&$\text{tr}[U(g)]$ in the order of Ref.~\cite{ITC}\\ \hline
$Y=(0,\frac{\pi}{b},0)$
&$(1,\xi_2,\bar{1},\bar{\xi}_2,1,\xi_2,\bar{1},\bar{\xi}_2)$\\
&$(1,\xi_2,\bar{1},\bar{\xi}_2,\bar{1},\bar{\xi}_2,1,\xi_2)$\\ \hline
$X=(\frac{\pi}{a},0,0)$
&$(1,\xi_1,\bar{\xi}_1,\bar{1},1,\xi_1,\bar{\xi}_1,\bar{1})$\\
&$(1,\xi_1,\bar{\xi}_1,\bar{1},\bar{1},\bar{\xi}_1,\xi_1,1)$\\ \hline
$Z=(0,0,\frac{\pi}{c})$&$(2,\bar{2},0,0,0,0,0,0)$\\\hline
$U=(\frac{\pi}{a},0,\frac{\pi}{c})$&$(2,2\xi_1,0,0,0,0,0,0)$\\\hline
$T=(0,\frac{\pi}{b},\frac{\pi}{c})$&$(2,2\xi_2,0,0,0,0,0,0)$\\\hline
$S=(\frac{\pi}{a},\frac{\pi}{b},0)$&$(1,\xi_3,\bar{\xi}_1,\bar{\xi}_2,1,\xi_3,\bar{\xi}_1,\bar{\xi}_2)$\\
&$(1,\xi_3,\bar{\xi}_1,\bar{\xi}_2,\bar{1},\bar{\xi}_3,\xi_1,\xi_2)$\\\hline
$R=(\frac{\pi}{a},\frac{\pi}{b},\frac{\pi}{c})$&$(2,2\xi_3,0,0,0,0,0,0)$\\\hline\hline
\end{tabular}
\end{center}
\end{table}

\begin{table}[h!]
\begin{center}
\caption{Solutions of the compatibility relations for \sg{50} ($Pban$).  $\xi_1^2=\xi_2^2=1$.\label{50}}
\begin{tabular}{c|c}\hline\hline
High-sym. momentum&$\text{tr}[U(g)]$ in the order of Ref.~\cite{ITC}\\ \hline
$Y=(0,\frac{\pi}{b},0)$&$(2,0,\bar{2},0,0,0,0,0)$\\\hline
$X=(\frac{\pi}{a},0,0)$&$(2,0,0,\bar{2},0,0,0,0)$\\\hline
$Z=(0,0,\frac{\pi}{c})$&$(1,\bar{1},\xi_1,\bar{\xi}_1,1,\bar{1},\xi_1,\bar{\xi}_1)$\\
&$(1,\bar{1},\xi_1,\bar{\xi}_1,\bar{1},1,\bar{\xi}_1,\xi_1)$\\
\hline
$U=(\frac{\pi}{a},0,\frac{\pi}{c})$&$(2,0,0,\bar{2}\xi_1,0,0,0,0)$\\\hline
$T=(0,\frac{\pi}{b},\frac{\pi}{c})$&$(2,0,2\xi_1,0,0,0,0,0)$\\\hline
$S=(\frac{\pi}{a},\frac{\pi}{b},0)$&$(2,2\xi_2,0,0,0,0,0,0)$\\\hline
$R=(\frac{\pi}{a},\frac{\pi}{b},\frac{\pi}{c})$&$(2,2\xi_2,0,0,0,0,0,0)$\\\hline\hline
\end{tabular}
\end{center}
\end{table}

\begin{table}[h!]
\begin{center}
\caption{Solutions of the compatibility relations for \sg{68} ($Ccce$). $*$ means that the corresponding operation is not a symmetry.\label{68}}
\begin{tabular}{c|c}\hline\hline
High-sym. momentum&$\text{tr}[U(g)]$ in the order of Ref.~\cite{ITC}\\ \hline
$Y=(0,\frac{2\pi}{b},0)$&$(1,\bar{1},\bar{1},1,1,\bar{1},\bar{1},1)$\\& $(1,\bar{1},\bar{1},1,\bar{1},1,1,\bar{1})$\\\hline
$Z=(0,0,\frac{\pi}{c})$&$(2,\bar{2},0,0,0,0,0,0)$\\\hline
$T=(0,\frac{2\pi}{b},\frac{\pi}{c})$&$(2,\bar{2},0,0,0,0,0,0)$\\\hline
$S=(\frac{\pi}{a},\frac{\pi}{b},0)$&$(2,0,*,*,0,0,*,*)$\\\hline
$R=(\frac{\pi}{a},\frac{\pi}{b},\frac{\pi}{c})$&$(2,0,*,*,0,0,*,*)$
\\\hline\hline
\end{tabular}
\end{center}
\end{table}

\begin{table}[h!]
\begin{center}
\caption{Solutions of the compatibility relations for \sg{86} ($P4_2/n$). $*$ means that the corresponding operation is not a symmetry.\label{86}}
\begin{tabular}{c|c}\hline\hline
High-sym. momentum&$\text{tr}[U(g)]$ in the order of Ref.~\cite{ITC}\\ \hline
$M=(\frac{\pi}{a},\frac{\pi}{a},0)$&$(2,2,0,0,0,0,0,0)$\\\hline
$Z=(0,0,\frac{\pi}{c})$&$(2,\bar{2},0,0,0,0,0,0)$\\\hline
$A=(\frac{\pi}{a},\frac{\pi}{a},\frac{\pi}{c})$&$(1,1,1,1,\xi,\xi,\xi,\xi)$\\&$(1,1,\bar{1},\bar{1},\bar{\xi},\bar{\xi},\xi,\xi)$\\\hline
$R=(0,\frac{\pi}{a},\frac{\pi}{c})$&$(2,0,*,*,0,0,*,*)$\\\hline
$X=(0,\frac{\pi}{a},0)$&$(2,0,*,*,0,0,*,*)$\\\hline
\hline
\end{tabular}
\end{center}
\end{table}

\begin{table*}[t]
\center
\caption{Solutions of the compatibility relations for \sg{191} ($P6/mmm$). In this table, $*$ means that the corresponding operation is not a symmetry. $\bar{1}$ is a shorthand for $-1$. \label{191}}
\begin{tabular}{c|c}
\hline \hline
High-sym. momentum& $\text{tr}[U(g)]$ in the order of Ref.~\cite{ITC} \\\hline
$K=(\frac{4\pi}{3},0,0)$,$K'=(\frac{2\pi}{3},\frac{2\pi}{\sqrt{3}},0)$ & $(2 ,2 ,2 ,* ,* ,* ,\bar{2} ,\bar{2} ,\bar{2} ,* ,* ,* ,* ,* ,* ,0 ,0 ,0 ,* ,* ,* ,0 ,0 ,0)$ \\
$M=(\pi,\frac{\pi}{\sqrt{3}},0)$ & $(2 ,* ,* ,2 ,* ,* ,* ,* ,\bar{2} ,* ,* ,\bar{2} ,0 ,* ,* ,0 ,* ,* ,* ,* ,0 ,* ,* ,0)$ \\
$M'=(0,\frac{2\pi}{\sqrt{3}},0)$ & $(2 ,* ,* ,2 ,* ,* ,* ,\bar{2} ,* ,* ,\bar{2} ,* ,0 ,* ,* ,0 ,* ,* ,* ,0 ,* ,* ,0 ,*)$ \\
$M''=(-\pi,\frac{\pi}{\sqrt{3}},0)$ & $(2 ,* ,* ,2 ,* ,* ,\bar{2} ,* ,* ,\bar{2} ,* ,* ,0 ,* ,* ,0 ,* ,* ,0 ,* ,* ,0 ,* ,*)$ \\
$A=(0,0,\pi)$ & $(2 ,\bar{1} ,\bar{1} ,\bar{2} ,1 ,1 ,0 ,0 ,0 ,0 ,0 ,0 ,2 ,\bar{1} ,\bar{1} ,\bar{2} ,1 ,1 ,0 ,0 ,0 ,0 ,0 ,0)$ \\
$H=(\frac{4\pi}{3},0,\pi)$,$H'=(\frac{2\pi}{3},\frac{2\pi}{\sqrt{3}},\pi)$ & $(2 ,2 ,2 ,* ,* ,* ,0 ,0 ,0 ,* ,* ,* ,* ,* ,* ,\bar{2} ,\bar{2} ,\bar{2} ,* ,* ,* ,0 ,0 ,0)$ \\
$L=(\pi,\frac{\pi}{\sqrt{3}},\pi)$ & $(2 ,* ,* ,2 ,* ,* ,* ,* ,0 ,* ,* ,0 ,\bar{2} ,* ,* ,\bar{2} ,* ,* ,* ,* ,0 ,* ,* ,0)$ \\
$L'=(0,\frac{2\pi}{\sqrt{3}},\pi)$ & $(2 ,* ,* ,2 ,* ,* ,* ,0 ,* ,* ,0 ,* ,\bar{2} ,* ,* ,\bar{2} ,* ,* ,* ,0 ,* ,* ,0 ,*)$ \\
$L''=(-\pi,\frac{\pi}{\sqrt{3}},\pi)$ & $(2 ,* ,* ,2 ,* ,* ,0 ,* ,* ,0 ,* ,* ,\bar{2} ,* ,* ,\bar{2} ,* ,* ,0 ,* ,* ,0 ,* ,*)$ \\
\hline \hline
\end{tabular}
\label{tab:summary}
\end{table*}

\begin{table*}
\begin{center}
\caption{
Solutions of the compatibility relations for the two gapless photonic bands of \sg{214} ($I4_132$) except for the singular $\Gamma$ point. In this table, $*$ means that the corresponding operation is not a symmetry. $\bar{1}$ is a shorthand for $-1$. \label{214}}
\begin{tabular}{c|c}\hline\hline
High-sym. momentum& $\text{tr}[U(g)]$ in the order of Ref.~\cite{ITC} \\\hline
$H=(0,\frac{2\pi}{a},0)$  &$(2,2,2,2,\bar{1},\bar{1},\bar{1},\bar{1},\bar{1},\bar{1},\bar{1},\bar{1},0,0,0,0,0,0,0,0,0,0,0,0)$\\\hline
$N=(\frac{\pi}{a},\frac{\pi}{a},0)$ &$(1,1,*,*,*,*,*,*,*,*,*,*,\bar{1},\bar{1},*,*,*,*,*,*,*,*,*,*)$\\&$(1,\bar{1},*,*,*,*,*,*,*,*,*,*,\bar{1},1,*,*,*,*,*,*,*,*,*,*)$\\\hline
$P=(\frac{\pi}{a},\frac{\pi}{a},\frac{\pi}{a})$& $(2,0,0,0,\bar{1},\bar{1},\bar{1},\bar{1},\bar{1},1,1,1,*,*,*,*,*,*,*,*,*,*,*,*)$\\\hline\hline
\end{tabular}
\end{center}
\end{table*}

\begin{table*}
\begin{center}
\caption{
Solutions of the compatibility relations for \sg{221} ($Pm\bar{3}m$). In this table, $*$ means that the corresponding operation is not a symmetry. $\bar{1}$ is a shorthand for $-1$. \label{221}}
\begin{tabular}{c|c}\hline\hline
High-sym. momentum& $\text{tr}[U(g)]$ in the order of Ref.~\cite{ITC} \\\hline
$X=(\frac{\pi}{a},0,0)$  &$(2,0,0,\bar{2},*,*,*,*,*,*,*,*,*,*,*,*,0,0,0,0,*,*,*,*,2\xi,\phantom{1}0,\phantom{1}0,\bar{2}\xi,*,*,*,*,*,*,*,*,*,*,*,*,0,0,0,0,*,*,*,*)$\\\hline
$M=(\frac{\pi}{a},\frac{\pi}{a},0)$ &
$(1,1,1,1,*,*,*,*,*,*,*,*,\bar{1},\bar{1},\bar{1},\bar{1},*,*,*,*,*,*,*,*,\phantom{1}\bar{\xi},\phantom{1}\bar{\xi},\phantom{1}\bar{\xi},\phantom{1}\bar{\xi},*,*,*,*,*,*,*,*,\xi,\xi,\xi,\xi,*,*,*,*,*,*,*,*)$\\&
$(1,1,\bar{1},\bar{1},*,*,*,*,*,*,*,*,\bar{1},\bar{1},1,1,*,*,*,*,*,*,*,*,\phantom{1}\xi,\phantom{1}\xi,\phantom{1}\bar{\xi},\phantom{1}\bar{\xi},*,*,*,*,*,*,*,*,\bar{\xi},\bar{\xi},\xi,\xi,*,*,*,*,*,*,*,*)$\\\hline
$R=(\frac{\pi}{a},\frac{\pi}{a},\frac{\pi}{a})$& $(2,2,2,2,\bar{1},\bar{1},\bar{1},\bar{1},\bar{1},\bar{1},\bar{1},\bar{1},0,0,0,0,0,0,0,0,0,0,0,0,\bar{2}\xi,\bar{2}\xi,\bar{2}\xi,\bar{2}\xi,\xi,\xi,\xi,\xi,\xi,\xi,\xi,\xi,0,0,0,0,0,0,0,0,0,0,0,0)$\\\hline\hline
\end{tabular}
\end{center}
\end{table*}

\end{document}